\documentclass[copyright,creativecommons]{eptcs}
\providecommand{\event}{Express/SOS 2025}

\usepackage{underscore}
\usepackage[T1]{fontenc}

\usepackage[english]{babel}
\usepackage[babel]{microtype}
\usepackage[all]{nowidow}
\usepackage{mathtools}
\usepackage{amssymb, amsthm, stmaryrd}
\usepackage{multirow, nicefrac, orcidlink, paralist}
\usepackage{tikz}
\usepackage{xcolor, xspace}
\usepackage{bookmark}
\usepackage{ftmpst-loop}

\title{Fault-Tolerant Multiparty Session Types\\ with Global Escape Loops}
\newcommand{\titlerunning}{FTMPST with Global Escape Loops}

\author{Lukas Bartl\,\orcidlink{0009-0000-2439-5025}
	\qquad\qquad Julian Linne\,\orcidlink{0009-0001-6040-597X}
	\qquad\qquad Kirstin Peters\,\orcidlink{0000-0002-4281-0074}
	\institute{Universität Augsburg, Germany}}
\newcommand{\authorrunning}{L. Bartl, J. Linne, \& K. Peters}

\hypersetup{colorlinks,
	pdftitle={Fault-Tolerant Multiparty Session Types with Global Escape Loops},
	pdfauthor={Lukas Bartl, Julian Linne, and Kirstin Peters},
	pdfkeywords={Multiparty Session Types, Fault-Tolerance, Distributed Algorithms}}

\begin{document}

\maketitle

%%%%%%%%%%%%%%
%  abstract  %
%%%%%%%%%%%%%%

\begin{abstract}
	Multiparty session types are designed to abstractly capture the structure of communication protocols and verify behavioural properties. One important such property is progress, \ie the absence of deadlock. Distributed algorithms often resemble multiparty communication protocols. But proving their properties, in particular termination that is closely related to progress, can be elaborate. Since distributed algorithms are often designed to cope with faults, a first step towards using session types to verify distributed algorithms is to integrate fault-tolerance.

	We extend FTMPST---a version of fault-tolerant multiparty session types with failure patterns to represent system requirements for system failures such as unreliable communication and process crashes---by a novel, fault-tolerant loop construct with global escapes that does not require global coordination.
	Each process runs its own local version of the loop.
	If a process finds a solution to the considered problem, it does not only terminate its own loop but also informs the other participants via $ \mathtt{exit} $-messages.
	Upon receiving an $ \mathtt{exit} $-message, a process immediately terminates its algorithm.
	To increase efficiency and model standard fault-tolerant algorithms, these messages are non-blocking, \ie a process may continue until a possibly delayed $ \mathtt{exit} $-message is received.
	To illustrate our approach, we analyse a variant of the well-known rotating coordinator algorithm by Chandra and Toueg.
\end{abstract}

%%%%%%%%%%%%%%%%%%
%  introduction  %
%%%%%%%%%%%%%%%%%%

\section{Introduction}
\label{sec:introduction}

Multi-Party Session Types (\MPST) are used to statically ensure correctly coordinated behaviour in systems without global control \cite{hondaYoshidaCarbone16,gentleintro}.
One important such property is progress, \ie the absence of deadlock. Like with every other static typing approach, the main advantage is their efficiency, \ie they avoid the problem of state space explosion.
\MPST are designed to abstractly capture the structure of communication protocols.
They describe global behaviours as \emph{sessions}, \ie units of
conversations \cite{hondaYoshidaCarbone16,BettiniAtall08,BocciAtall10}. The participants of such sessions are called \emph{roles}.
\emph{Global types} specify protocols from a global point of view.
These types are used to reason about processes formulated
in a \emph{session calculus}.

Distributed algorithms (DA) very much resemble multiparty communication protocols.
An essential behavioural property of DA is termination \cite{Lynch96,Tel2000}, despite failures, but it is often elaborate to prove.
It turns out that progress (as provided by MPST) and termination (as required by DA) are closely related.

Many DA were designed in a fault-tolerant way, in order to work in environments, where they have to cope with system failures---be it links dropping messages or processes crashing.
We focus on masking fault-tolerant algorithms (see \cite{DBLP:journals/csur/Gartner99}), \ie safety and liveness requirements hold despite failures without further intervention by the programmer.

While the detection of conceptual design errors is a standard property of type systems, proving correctness of algorithms despite the occurrence of system failures is not.
Likewise, traditional \MPST do not cover fault tolerance or failure handling.
There are several approaches to integrate explicit failure handling in \MPST (\eg \cite{CarboneHondaYoshida08,capecchi2016,chen.viering.bejleri.ziarek.eugster,viering18,dem15,adameitPetersNestmann17}).
These approaches are sometimes enhanced with recovery mechanisms such as \cite{castellaniEtAl} or even provide algorithms to help find safe states to recover from, as in \cite{neykova2017let}.
Many of these approaches introduce nested \textsc{try}-and-\textsc{catch}-blocks and a challenge is to ensure that all participants are consistently informed about concurrent \textsc{throws} of exceptions.
Therefore, exceptions are propagated within the system.
Though explicit failure handling makes sense for high-level applications, the required message overhead is too inefficient for many low-level algorithms.
Instead, these low-level algorithms are often designed to tolerate a certain amount of failures.
Since we focus on the communication structure of systems, additional messages as reaction to faults (\eg to propagate faults) are considered \emph{non-masking} failure handling.
In contrast, we expect masking fault-tolerant algorithms to cope without messages triggered by faults.
We study how much unhandled failures a well-typed system can tolerate, while maintaining the typical properties of \MPST.

Type systems are usually designed for failure-free scenarios.
An exception is \cite{KouzapasGutkovasGay14} that introduces unreliable broadcast, where a transmission can be received by multiple receivers but not necessarily all available receivers. In the latter case, the receiver is deadlocked. In contrast, we consider fault-tolerant interactions, where in the case of a failure the receiver is \emph{not} deadlocked.

The already mentioned systems in \cite{CarboneHondaYoshida08,capecchi2016,chen.viering.bejleri.ziarek.eugster,viering18,dem15} extend session types with exceptions thrown by processes within \textsc{try}-and-\textsc{catch}-blocks, interrupts, or similar syntax.
They structurally and semantically encapsulate an unreliable part of a protocol and provide some means to 'detect' a failure and 'react' to it.
Here we deliberately do not model how to 'detect' a failure.
Different system architectures might provide different mechanisms to do so, for example, by means of time-outs.
As is standard for the analysis of DA, our approach allows us to port the verified algorithms on different system architectures that satisfy the necessary system requirements.

Another essential difference is how systems react to faults.
In~\cite{capecchi2016}, \textsc{throw}-messages are propagated among nested \textsc{try}-and-\textsc{catch}-blocks to ensure that all participants are \emph{consistently} informed about concurrent \textsc{throws} of exceptions.
Fault-tolerant DA, however, have to deal with the problem of inconsistency; one of their most challenging problems.
\emph{Distributed} processes usually cannot reliably observe an error on another system part, unless they are informed by some
system ``device'' (like the ``coordinator'' of~\cite{viering18} or the ``oracle'' of~\cite{capecchi2016}).
Therefore, abstractions like unreliable failure detectors are used to model this restricted observability which can, for example, be implemented by time-outs.

\looseness=-1
We extend FTMPST \cite{petersNestmannWagner22, petersNestmannWagner23}, a version of fault-tolerant multiparty session types with failure patterns to represent system requirements for system failures such as unreliable communication and process crashes.
We add a novel, fault-tolerant loop construct with global escapes but without a need for global coordination.
Thereby, we tackle an open question of \cite{petersNestmannWagner23}, namely how to conveniently type unreliable recursive parts of protocols.
Distributed algorithms are often recursive and exit this recursion if a result was successfully computed.
In \cite{petersNestmannWagner23}, \weakR branching was used to exit a standard recursion.
Unfortunately, this operation temporarily blocks some processes.
Our novel loop construct overcomes this problem.

Each loop of an algorithm has a unique identifier, where unique means from a global point of view.
Each process runs its own local version of the loop, but the local loops that jointly define a recursive routine of the algorithm have the same identifier.
If a process finds a solution to the considered problem, it does not only terminate its own loop but also informs the other participants via $ \mathtt{exit} $-messages that may carry a solution value.
Upon receiving an $ \mathtt{exit} $-message, a process immediately terminates its algorithm.
To increase efficiency and model standard fault-tolerant algorithms, these messages are non-blocking, \ie a process may continue until a possibly delayed $ \mathtt{exit} $-message is received.
Since communication in the system is asynchronous and because of faults such as message delays, many algorithms do not forbid that different participants terminate the protocol concurrently.
Hence, there may be several concurrent $ \mathtt{exit} $-messages for the same local loop.
The algorithm then has to ensure, that all of them carry the same solution value---usually called \emph{agreement}.

To guide the behaviour of unreliable communication, we inherit from \cite{petersNestmannWagner23} the \emph{failure patterns} used in the semantics of processes.
Note that these patterns are not defined, but \emph{could} be instantiated by an application.
This allows us to cover requirements on the system---as, \eg, a bound on the number of faulty processes---as well as more abstract concepts like failure detectors.
It is beyond the scope of this paper to discuss \emph{how} failure patterns could be implemented.
To illustrate our approach we analyse a variant of the well-known rotating coordinator algorithm by Chandra and Toueg.

Additional material and the missing proofs are contained in a technical report \cite{bartlLinnePetersTecRep25}.

%%%%%%%%%%%%%%%%%%%%%%%%%%%%%%%%%%%%%%%%
%  Fault-Tolerant Types and Processes  %
%%%%%%%%%%%%%%%%%%%%%%%%%%%%%%%%%%%%%%%%

\section{Fault-Tolerant Types and Processes}
\label{sec:syntax}

Following \cite{petersNestmannWagner23}, we consider three levels of failures in interactions:
\begin{compactdesc}
	\item[\StrongR ($ \iR $)] Neither the sender nor the receiver can crash as long as they are involved in this interaction. The message cannot be lost by the communication medium. This form corresponds to reliable communication as it was described in \cite{aguileraChenToueg97} in the context of distributed algorithms.
		This is the standard, failure-free case.
	\item[\WeakR ($ \iW $)] Both the sender and the receiver might crash at every possible point during this interaction. But the communication medium cannot lose the message.
	\item[\Unrel ($ \iU $)] Both the sender and the receiver might crash at every possible point during this interaction and the communication medium might lose the message. There are no guarantees that this interaction---or any part of it---takes place.
		Here, it is difficult to ensure interesting properties in branching.
\end{compactdesc}
We use the subscripts or superscripts $ \iR $, $ \iW $, or $ \iU $ to indicate actions of the respective kind.
Our new loop construct relies on \unrel interactions for the loop body such that the termination of the loop does not cause any blocking of the interaction partners.
However, the $ \mathtt{exit} $-messages should not be dropped before the loop is terminated and are thus \weakR.

For clarity, we often distinguish names into \emph{values}, \ie the payload of messages, \emph{shared channels}, or \emph{session channels} according to their usage; there is, however, no need to formally distinguish between different kinds of names.

We assume that the sets $ \names $ of names $ \Chan[a], \Chan, \Args \ldots $; $ \roles $ of roles $ \Role[n], \Role, \ldots $; $ \labels $ of labels $ \Label, \LabelD, \ldots $; $ \typeVars $ of type variables $ \TypeV $; and $ \procVars $ of process variables $ \ProcV $ are pairwise distinct.
To simplify the reduction semantics of our session calculus, we use natural numbers as roles (compare to \cite{hondaYoshidaCarbone16}).
Sorts $ \Sort $ range over $ \mathbb{B}, \mathbb{N}, \ldots $.
The set $ \expressions $ of expressions $ \Expr, \Expr[v], \Expr[b], \ldots $ is constructed from the standard Boolean operations, natural numbers, standard arithmetic operators, tuples, names, and (in)equalities.
We assume an evaluation function $ \Eval{\cdot} $ that evaluates expressions to values.

Global types specify the desired communication structure from a global point of view.
In local types, this global view is projected to the specification of a single role/participant.
We start from standard \MPST \cite{hondaYoshidaCarbone08,hondaYoshidaCarbone16} extended by \unrel communication and \weakR branching in \cite{petersNestmannWagner22, petersNestmannWagner23}.
We then add an \textcolor{blue}{unreliable loop construct with \weakR global escapes} (highlighted in blue) in Figure~\ref{fig:syntax}.

\begin{figure}[t]
	\centering
	\renewcommand{\tabcolsep}{0pt}
	\begin{tabular}{|@{\,}rcl@{\,}|@{\,}rcl@{\,}|@{\,}rcl@{\,}|}
		\hline
		&&&&&&&& \\[-1em]
		\multicolumn{3}{|c|@{\,}}{Global Types} & \multicolumn{3}{c|@{\,}}{Local Types} & \multicolumn{3}{c|}{Processes}\\
		&&&&&& $ \PT $ & $ \deffTerms $ & $ \PReq{\Chan[a]}{\Role[n]}{\Chan}{P} \sepTerms \PAcc{\Chan[a]}{\Role}{\Chan}{P} $\\
		\multirow{2}{*}{$ \GT $} & \multirow{2}{*}{$ \deffTerms $} & \multirow{2}{*}{$ \GComR{\Role_1}{\Role_2}{\Sort}{\GT} $} & $ \LT $ & $ \deffTerms $ & $ \LSendR{\Role_2}{\Sort}{\LT} $ && $ \sepTerms $ & $ \PSendR{\Chan}{\Role_1}{\Role_2}{\Expr}{P} $\\
		&&&& $ \sepTerms $ & $ \LGetR{\Role_1}{\Sort}{\LT} $ && $ \sepTerms $ & $ \PGetR{\Chan}{\Role_2}{\Role_1}{\Args}{\PT} $\\
		& \multirow{2}{*}{$ \sepTerms $} & \multirow{2}{*}{$ \GComU{\Role_1}{\Role_2}{\Label}{\Sort}{\GT} $} && $ \sepTerms $ & $ \LSendU{\Role_2}{\Label}{\Sort}{\LT} $ && $ \sepTerms $ & $ \PSendU{\Chan}{\Role_1}{\Role_2}{\Label}{\Expr}{P} $\\
		&&&& $ \sepTerms $ & $ \LGetU{\Role_1}{\Label}{\Sort}{\LT} $ && $ \sepTerms $ & $ \PGetU{\Chan}{\Role_2}{\Role_1}{\Label}{\Expr[v]}{\Args}{P} $\\
		& \multirow{2}{*}{$ \sepTerms $} & \multirow{2}{*}{$ \GBranR{\Role_1}{\Role_2}{\Set{ \Label_i.\GT_i}_{i \in \indexSet}} $} && $ \sepTerms $ & $ \LSelR{\Role_2}{\Set{ \Label_i.\LT_i }_{i \in \indexSet}} $ && $ \sepTerms $ & $ \PSelR{\Chan}{\Role_1}{\Role_2}{\Label}{P} $\\
		&&&& $ \sepTerms $ & $ \LBranR{\Role_1}{\Set{ \Label_i.\LT_i }_{i \in \indexSet}} $ && $ \sepTerms $ & $ \PBranR{\Chan}{\Role_2}{\Role_1}{\Set{ \Label_i.P_i }_{i \in \indexSet}} $\\
		& \multirow{2}{*}{$ \sepTerms $} & \multirow{2}{*}{$ \GBranW{\Role}{\Role[R]}{\Set{ \Label_i.\GT_i }_{i \in \indexSet, \LabelD}} $} && $ \sepTerms $ & $ \LSelW{\Role[R]}{\Set{ \Label_i.\LT_i }_{i \in \indexSet}} $ && $ \sepTerms $ & $ {\PSelW{\Chan}{\Role}{\Role[R]}{\Label}{P}} $\\
		&&&& $ \sepTerms $ & $ \LBranW{\Role}{\Set{ \Label_i.\LT_i }_{i \in \indexSet, \LabelD}} $ && $ \sepTerms $ & $ \PBranW{\Chan}{\Role_j}{\Role}{\Set{ \Label_i.P_i }_{i \in \indexSet, \LabelD}} $\\
		& $ \sepTerms $ & $ \GPar{\GT_1}{\GT_2} $ &&&&& $ \sepTerms $ & $ P_1 \mid P_2 $\\
		& $ \sepTerms $ & $ \GRep{\TypeV, \textcolor{blue}{\Counter}}{\GT} \sepTerms \TypeV $ && $ \sepTerms $ & $ \LRep{\TypeV, \textcolor{blue}{\Counter = n}}{\LT} \sepTerms \TypeV $ && $ \sepTerms $ & $ \PRep{\ProcV, \textcolor{blue}{\Counter = n}}{P} \sepTerms \ProcV $\\
		& $ \sepTerms $ & $ \GEnd $ && $ \sepTerms $ & $ \LEnd $ && $ \sepTerms $ & $ \PEnd $\\
		& $ \sepTerms $ & $ \textcolor{blue}{\GLoop{\Role[R]}{\Expr}{\Counter}{\Sort_0}{\GT_0}{\Sort_2}{\GT_2}} $ && $ \sepTerms $ & $ \textcolor{blue}{\LLoop{\Role[R]}{\Expr}{\Counter}{n}{\Sort_0}{\LT_0}{\LT_1}{\Sort_2}{\LT_2}} $ && $ \sepTerms $ & $ \textcolor{blue}{\PLoop{\PLoopHeader{\Chan}{\Role}{\Role[R]}{\Expr}{\Counter}{n}}{\Args}{\PT_0}{\PT_1}{\Args[y]}{\PT_2}} $\\
		& $ \sepTerms $ & $ \textcolor{blue}{\TCall{\Expr}} $ && $ \sepTerms $ & $ \textcolor{blue}{\TCall{\Expr}} $ && $ \sepTerms $ & $ \textcolor{blue}{\PCall{\Expr}{\Expr'}} \sepTerms \textcolor{blue}{\Exit{\Expr}{\Expr'}} $\\
		&&&&&&& $ \sepTerms $ & $ \PITE{\Expr[b]}{P_1}{P_2} $\\
		&&&&&&& $ \sepTerms $ & $ \PRes{\Args}{P} \sepTerms \PCrash $\\
		& \multirow{2}{*}{$ \sepTerms $} & \multirow{2}{*}{$ \GDel{\Role_1}{\Role_2}{\Chan'}{\Role}{\LT}{\GT} $} && $ \sepTerms $ & $ \LDelA{\Role_2}{\Chan'}{\Role}{\LT}{\LT'} $ && $ \sepTerms $ & $ \PDelA{\Chan}{\Role_1}{\Role_2}{\AT{\Chan'}{\Role}}{\PT} $\\
		&&&& $ \sepTerms $ & $ \LDelB{\Role_1}{\Chan'}{\Role}{\LT}{\LT'} $ && $ \sepTerms $ & $ \PDelB{\Chan}{\Role_2}{\Role_1}{\AT{\Chan'}{\Role}}{\PT} $\\
		&&&&&&& $ \sepTerms $ & $ \MQ{\Chan}{\Role_1}{\Role_2}{\Queue} $\\
		\hline
		&& \multicolumn{3}{l}{} &&&& \\[-1em]
		\multicolumn{6}{|c|@{\,}}{Message Types} & \multicolumn{3}{c|}{Messages}\\
		$ \mathsf{mt} $ & $ \deffTerms $ & \multicolumn{3}{l}{$ \MessR{\Sort} \sepTerms \MessU{\Label}{\Sort} \sepTerms \MessBR{\Label} \sepTerms \MessBW{\Label} $} && $ \mathsf{m} $ & $ \deffTerms $ & $ \MessR{\Expr[v]} \sepTerms \MessU{\Label}{\Expr[v]} \sepTerms \MessBR{\Label} \sepTerms \MessBW{\Label} $\\
		& $ \sepTerms $ & \multicolumn{3}{l}{$ \textcolor{blue}{\Exit{\Expr[id]}{\Sort}} \sepTerms \AT{\Chan}{\Role} $} &&& $ \sepTerms $ & $ \textcolor{blue}{\Exit{\Expr[id]}{\Expr[v]}} \sepTerms \AT{\Chan}{\Role} $\\
		$ \MT $ & $ \deffTerms $ & \multicolumn{3}{l}{$ \emptyList \sepTerms \mathsf{mt} \# \MT $} && $ \Queue $ & $ \deffTerms $ & $ \emptyList \sepTerms \mathsf{m} \# \Queue $\\
		\hline
	\end{tabular}
	\caption{Syntax of Fault-Tolerant \MPST with \textcolor{blue}{Global Escape Loops.}}
	\label{fig:syntax}
\end{figure}

A new session $ \Chan $ with $ \Role[n] $ roles is initialised with $ \PReq{\Chan[a]}{\Role[n]}{\Chan}{P} $ and $ \PAcc{\Chan[a]}{\Role}{\Chan}{P} $ via the shared channel $ \Chan[a] $. We identify sessions with their unique session channel.

The type $ \GComR{\Role_1}{\Role_2}{\Sort}{\GT} $ specifies a \strongR communication from role $ \Role_1 $ to role $ \Role_2 $ to transmit a value of sort $ \Sort $ and then continues with $ \GT $.
A system with this type will be guaranteed to perform a corresponding action.
In a session $ \Chan $ this communication is implemented by the sender $ \PSendR{\Chan}{\Role_1}{\Role_2}{\Expr}{\PT_1} $ (specified as $ \LSendR{\Role_2}{\Sort}{\LT_1} $) and the receiver $ \PGetR{\Chan}{\Role_2}{\Role_1}{\Args}{\PT_2} $ (specified as $ \LGetR{\Role_1}{\Sort}{\LT_2} $).
As a result, the receiver instantiates $ \Args $ in its continuation $ \PT_2 $ with the received value.

The type $ \GComU{\Role_1}{\Role_2}{\Label}{\Sort}{\GT} $ specifies an \unrel communication from $ \Role_1 $ to $ \Role_2 $ transmitting (if successful) a label $ \Label $ and a value of sort $ \Sort $ and then continues (regardless of the success of this communication) with $ \GT $.
The \unrel counterparts of senders and receivers are $ \PSendU{\Chan}{\Role_1}{\Role_2}{\Label}{\Expr}{\PT_1} $ (specified as $ \LSendU{\Role_2}{\Label}{\Sort}{\LT_1} $) and $ \PGetU{\Chan}{\Role_2}{\Role_1}{\Label}{\Args[v]}{\Args}{\PT_2} $ (specified as $ \LGetU{\Role_1}{\Label}{\Sort}{\LT_2} $).
The receiver $ \PGetU{\Chan}{\Role_2}{\Role_1}{\Label}{\Args[v]}{\Args}{\PT_2} $ declares a default value $ \Args[v] $ that is used instead of a received value to instantiate $ \Args $ after a failure.
Moreover, a label is communicated that helps us to ensure that a faulty \unrel communication does not influence later actions.

The \strongR branching $ \GBranR{\Role_1}{\Role_2}{\Set{ \Label_i.\GT_i}_{i \in \indexSet}} $ allows $ \Role_1 $ to pick one of the branches offered by $ \Role_2 $.
We identify the branches with their respective label.
Selection of a branch is by $ \PSelR{\Chan}{\Role_1}{\Role_2}{\Label}{P} $ (specified as $ \LSelR{\Role_2}{\Set{ \Label_i.\LT_i }_{i \in \indexSet}} $).
Upon receiving $ \Label_j $, $ \PBranR{\Chan}{\Role_2}{\Role_1}{\Set{ \Label_i.P_i }_{i \in \indexSet}} $ (specified as $ \LBranR{\Role_1}{\Set{ \Label_i.\LT_i }_{i \in \indexSet}} $) continues with $ \PT_j $.

As discussed in \cite{petersNestmannWagner23}, the counterpart of branching is \weakR and not \unrel.
It is implemented by $ \GBranW{\Role}{\Role[R]}{\Set{\Label_i.\GT_i}_{i \in \indexSet, \LabelD}} $, where $ \Role[R] \subseteq \roles $ and $ \LabelD $ with $ \default \in \indexSet $ is the default branch.
We use a broadcast from $ \Role $ to all roles in $ \Role[R] $ to ensure that the sender can influence several participants consistently (see \cite{petersNestmannWagner23} for an explanation).
The type system ensures that all processes that are not crashed will move to the same branch.
We often abbreviate branching \wrt a small set of branches by omitting the set brackets and instead separating the branches by $ \oplus $, where the last branch is always the default branch.
In contrast to the \strongR cases, $ \PSelW{\Chan}{\Role}{\Role[R]}{\Label}{\PT} $ (specified as $ \LSelW{\Role[R]}{\Set{ \Label_i.\LT_i }_{i \in \indexSet}} $) allows to broadcast its decision to $ \Role[R] $ and $ \PBranW{\Chan}{\Role_j}{\Role}{\Set{ \Label_i.\PT_i }_{i \in \indexSet, \LabelD}} $ (specified as $ \LBranW{\Role}{\Set{ \Label_i.\LT_i }_{i \in \indexSet, \LabelD}} $) defines a default label $ \LabelD $.

We extend the standard operators for recursion $ \GRep{\TypeV}{\GT} $, $ \LRep{\TypeV}{\LT} $, and $ \PRep{\ProcV}{\PT} $ of \cite{petersNestmannWagner23} by a counter $ \PRep{\ProcV, \textcolor{blue}{\Counter = n}}{\PT} $ (specified as $ \LRep{\TypeV, \textcolor{blue}{\Counter = n}}{\LT} $ with the global type $ \GRep{\TypeV, \textcolor{blue}{\Counter}}{\GT} $), where $ n $ is a natural number that is increased by unfolding recursion and $ \Counter $ can be used as pointer to the current value of the counter within expressions in $ \GT $, $ \LT $, and $ \PT $.
These expressions allow us to construct unique identifiers for loops within a surrounding recursion.

\looseness=-1
A loop $ \textcolor{blue}{\PLoop{\PLoopHeader{\Chan}{\Role}{\Role[R]}{\Expr}{\Counter}{n}}{\Args}{\PT_0}{\PT_1}{\Args[y]}{\PT_2}} $ (specified as $ \textcolor{blue}{\LLoop{\Role[R]}{\Expr}{\Counter}{n}{\Sort_0}{\LT_0}{\LT_1}{\Sort_2}{\LT_2}} $ with the global type $ \textcolor{blue}{\GLoop{\Role[R]}{\Expr}{\Counter}{\Sort_0}{\GT_0}{\Sort_2}{\GT_2}} $) creates a loop in that role $ \Role $ of session $ \Chan $ is currently running the \emph{loop body} $ P_1 $ and may interact with the roles in $ \Role[R] $ that are running their local versions of this loop.
We identify a loop with its unique identifier $ \Expr[id] = \Eval{\Expr} $ that is unique for the whole derivation of the system and the same for all roles $ \Role[R] \cup \Set{\Role} $.
Again, the loop has a counter \textcolor{blue}{$ \Counter = n $} that is increased in unfolding loops and can be used to create the unique identifiers of loops nested within the current loop.
Communication within a loop is \unrel.
With $ \textcolor{blue}{\PCall{\Expr[id]}{\Expr[v]}} $ (specified as \textcolor{blue}{$ \TCall{\Expr[id]} $}) role $ \Role $ invokes another iteration of the loop $ \Expr[id] $ given by the \emph{loop program} $ {\left( x \right)}.P_0 $, where $ x $ is instantiated with $ \Expr[v] $.
Role $ \Role $ can terminate its own loop and all loops of the other $ \Role[R] $ by sending $ \textcolor{blue}{\Exit{\Expr[id]}{\Expr[v]'}} $.
In this case, or upon receiving $ \textcolor{blue}{\Exit{\Expr[id]}{\Expr[v]'}} $, role $ \Role $ continues with the \emph{loop continuation} $ {\left( y \right)}.P_2 $ of loop $ \Expr[id] = \Eval{\Expr} $, where $ y $ is instantiated by $ \Expr[v]' $.
The loop body $ P_1 $ contains whatever is left of the current iteration of the loop program $ P_0 $.
We initialise, as expected by the type system, a loop as $ \textcolor{blue}{\PLoop{\PLoopHeader{\Chan}{\Role}{\Role[R]}{\Expr}{\Counter}{0}}{\Args}{\PT_0}{\PCall{\Expr}{\Expr[v]}}{\Args[y]}{\PT_2}} $ such that its first step calls the first iteration of the loop.

The $ \PCrash $ denotes a process that crashed.
Similar to \cite{hondaYoshidaCarbone16}, we use message queues to implement asynchrony in sessions.
Therefore, session initialisation introduces a directed and initially empty message queue $ \MQ{\Chan[s]}{\Role_i}{\Role_j}{\emptyList} $ for each pair of roles $ \Role_i \neq \Role_j $ of the session $ \Chan[s] $.
The separate message queues ensure that messages with different sources or destinations are not ordered, but each message queue is FIFO.
Since the different forms of interaction might be implemented differently (\eg by TCP or UDP), it makes sense to further split the message queues into three message queues for each pair $ \Role_i \neq \Role_j $ such that different kinds of messages do not need to be ordered.
To simplify the presentation of examples in this paper and not to blow up the number of message queues, we stick to a single message queue for each pair $ \Role_i \neq \Role_j $.
However, the correctness of our type system does not depend on this decision.
We have six kinds of messages $ \mathsf{m} $ and corresponding message types $ \mathsf{mt} $ in Figure~\ref{fig:syntax}---one for each kind of interaction.
In \strongR communication, a value $ \Expr[v] $ (of sort $ \Sort $) is transmitted in a message $ \MessR{\Expr[v]} $ of type $ \MessR{\Sort} $.
In \unrel communication, the message $ \MessU{\Label}{\Expr[v]} $ (of type $ \MessU{\Label}{\Sort} $) additionally carries a label $ \Label $.
For branching, only the picked label $ \Label $ is transmitted and we add the kind of branching as superscript, \ie message/type $ \MessBR{\Label} $ is for \strongR branching and message/type $ \MessBW{\Label} $ for \weakR branching.
The message $ \textcolor{blue}{\Exit{\Expr[id]}{\Expr[v]}} $ of type $ \textcolor{blue}{\Exit{\Expr[id]}{\Sort}} $ signals that the loop $ \Expr[id] $ can be terminated.
Finally, the message/type $ \AT{\Chan}{\Role} $ is for session delegation.
A message queue $ \Queue $ is a list of messages $ \mathsf{m} $ and $ \MT $ is a list of message types $ \mathsf{mt} $.

The remaining operators for independence $ \GPar{\GT}{\GT'} $; parallel composition $ \PPar{\PT}{\PT'} $; inaction $ \GEnd $, $ \PEnd $; conditionals $ \PITE{\Expr[b]}{\PT_1}{\PT_2} $; session delegation $ \GDel{\Role_1}{\Role_2}{\Chan'}{\Role}{\LT}{\GT} $, $ \PDelA{\Chan}{\Role_1}{\Role_2}{\AT{\Chan'}{\Role}}{\PT} $, $ \PDelB{\Chan}{\Role_2}{\Role_1}{\AT{\Chan'}{\Role}}{\PT} $; and restriction $ \PRes{\Args}{\PT} $ are all standard.

As usual, we assume that recursion variables are guarded and do not occur free in types or processes and, similarly, that recursive calls $ \textcolor{blue}{\PCall{\Expr}{\Expr'}, \TCall{\Expr}} $ are guarded within loop programs and do not occur outside of the declaration of loop $ \Eval{\Expr} $ in types or processes.
To ensure that loops are uniquely identified, their identifiers are described as expressions that have to evaluate to a unique identifier in a type and all its unfoldings of recursion.
That is to say, within standard recursion or surrounding loops, these identifiers have to be built by a mechanism that ensures uniqueness, such as the counter of the surrounding recursion.
More precisely, all iterations of a loop have the same identifier, whereas a loop within a surrounding recursion or loop needs a fresh identifier for every iteration of the surrounding recursion or loop.
Moreover, the type system ensures that neither loop bodies nor loop programs may contain free type variables.

In types $ \GRep{\TypeV, \textcolor{blue}{\Counter}}{\GT} $ and $ \LRep{\TypeV, \textcolor{blue}{\Counter = n}}{\LT} $ the type variable $ \TypeV $ and the variable $ \Counter $ are \emph{bound} in $ \GT $, $ \LT $.
In processes $ \PRep{\ProcV, \textcolor{blue}{\Counter = n}}{\PT} $ the process variable $ \ProcV $ and the variable $ \Counter $ are bound in $ \PT $.
Similarly, in loops $ \textcolor{blue}{\GLoop{\Role[R]}{\Expr}{\Counter}{\Sort_0}{\GT_0}{\Sort_2}{\GT_2}} $, $ \textcolor{blue}{\LLoop{\Role[R]}{\Expr}{\Counter}{n}{\Sort_0}{\LT_0}{\LT_1}{\Sort_2}{\LT_2}} $, and $ \textcolor{blue}{\PLoop{\PLoopHeader{\Chan}{\Role}{\Role[R]}{\Expr}{\Counter}{n}}{\Args}{\PT_0}{\PT_1}{\Args[y]}{\PT_2}} $ the variable $ \Counter $ is bound in $ \GT_0 $, $ \LT_0 $, and $ \PT_0 $.
Additionally, all names in round brackets are bound in the remainder of the respective process, \eg $ \Chan $ is bound in $ \PT $ by $ \PReq{\Chan[a]}{\Role[n]}{\Chan}{\PT} $ and $ \Args $ is bound in $ \PT $ by $ \PGetR{\Chan}{\Role_1}{\Role_2}{\Args}{\PT} $.
A variable or name is \emph{free} if it is not bound. Let $ \FreeNames{\PT} $ return the free names of $ \PT $.

Let \emph{subterm} denote a (type or process) expression that syntactically occurs within another (type or process) term.
We use '$ . $' (as \eg in $ \PReq{\Chan[a]}{\Role}{\Chan}{\PT} $) to denote sequential composition.
In all operators the \emph{prefix} before '$ . $' guards the \emph{continuation} after the '$ . $'.
Moreover, a loop is a guard for its loop continuation, but its loop body is unguarded.
Let $ \prod_{1 \leq i \leq n} \PT_i $ abbreviate $ \PPar{\PT_1}{\PPar{\ldots}{\PT_n}} $.

Let $ \Roles{\GT} $ return all roles that occur in $ \GT $.
We write $ \NotStronglyReliable{\GT} $, $ \NotStronglyReliable{\LT} $, and $ \NotStronglyReliable{\PT} $, if none of the prefixes in $ \GT $, $ \LT $, and $ \PT $ is \strongR or for delegation and if $ \PT $, $ \GT $, or $ \LT $ do not contain message queues.
We write $ \Unreliable{A} $ if $ \NotStronglyReliable{A} $ and none of the prefixes in $ A $ is a \weakR branching.

A session channel and a role together uniquely identify a participant of a session, called an \emph{actor}. A process has an actor $ \AT{\Chan[s]}{\Role} $ if it has an action prefix or a loop on $ \Chan $ that mentions $ \Role $ as its first role.
Let $ \Actors{\PT} $ be the set of actors of $ \PT $.

As discussed in \cite{petersNestmannWagner23}, labels may carry additional runtime information such as time\-stamps, in order to provide the technical means to implement the failure patterns introduced with the semantics below.

Allowing for runtime information in labels requires a subtle difference in the way labels are used.
A timestamp may be added by the sender to capture the transmission time, but for the receiver it is hard to have this information already present in its label before or during reception.
Similarly, types in our static type system should not depend on any runtime information.
Hence, in contrast to standard \MPST, we do not expect the labels of senders and receivers as well as the labels of processes and types to match exactly.
Instead we assume a predicate $ \compL[] $ that compares two labels and is satisfied if the parts of the labels that do not refer to runtime information correspond.
If labels do not contain runtime information, $ \compL[] $ can be instantiated with equality.
We require that $ \compL[] $ is unambiguous on labels used in types, \ie given two labels of processes $ \Label_{\PT}, \Label_{\PT}' $ and two labels of types $ \Label_{\LT}, \Label_{\LT}' $ then $ \Label_{\PT} \compL \Label_{\PT}' \wedge \Label_{\PT} \compL \Label_{\LT} \Rightarrow \Label_{\PT}' \compL \Label_{\LT} $ and $ \Label_{\PT} \compL \Label_{\LT} \wedge \Label_{\LT} \nCompL \Label_{\LT}' \Rightarrow \Label_{\PT} \nCompL \Label_{\LT}' $.

Of course, the presented type system remains valid if we use labels without additional runtime information.
Interestingly, also the static information in labels, that have to coincide for senders and receivers and their types, can be exploited to guide communication.
In contrast to standard \MPST and to support \unrel communication, our \MPST variant will ensure that all occurrences of the same label are associated with the same sort.
This helps us in the case of failures to ensure the absence of communication mismatches, \ie the type of a transmitted value has to be the type that the receiver expects.
Similarly, labels are used in \cite{CairesVieira2010} to avoid communication errors.

Our type system verifies processes, \ie implementations, against a specification that is a global type.
Since processes implement local views, local types are used as a mediator between the global specification and the respective local end points.
To ensure that the local types correspond to the global type, they are derived by \emph{projection}.

Projection maps global types onto the respective local type for a given role $ \Role[p] $.
Recursion and loops are projected as follows:
\begin{align*}
	\Proj{\left( \GRep{\TypeV, \textcolor{blue}{\Counter}}{\GT} \right)}{\Role[p]} &\deff
		\begin{cases}
			\Proj{\GT \textcolor{blue}{\Subst{0}{\Counter}}}{\Role[p]} & \text{if } t \text{ does not occur in } \GT\\
			\LRep{\TypeV, \textcolor{blue}{\Counter = 0}}{\Proj{\GT}{\Role[p]}} & \text{else if } \Role[p] \in \Roles{\GT}\\
			\LEnd & \text{otherwise}
		\end{cases}
	\\
	\textcolor{blue}{\Proj{\left( \GLoop{\Role[R]}{\Expr}{\Counter}{\Sort_0}{\GT_0}{\Sort_2}{\GT_2} \right)}{\Role[p]}} & \textcolor{blue}{\deff
	\begin{cases}
		\LLoop{\Role[R] \setminus \Set{\Role[p]}}{\Expr}{\Counter}{0}{\Sort_0}{\Proj{\GT_0}{\Role[p]}}{\TCall{\Expr}}{\Sort_2}{\Proj{\GT_2}{\Role[p]}} & \text{if } \Role[p] \in \Role[R] \\
		\Proj{\GT_2}{\Role[p]} & \text{otherwise}
	\end{cases}}
\end{align*}
Projection of recursion is standard except for the initialisation of the counter \textcolor{blue}{$ \Counter $} with $ 0 $.
Recursive types without their recursion variable are mapped to the projection of their recursion body (similar to \cite{CastellaniAtAll20}), else if $ \Role[p] $ occurs in the recursion body we map to a recursive local type, or else to successful termination.
If projected on one of its roles $ \Role[p] \in \Role[R] $, the global specification of the loop program $ G_0 $ and the global specification of the loop continuation $ G_2 $ are projected on $ \Role[p] $.
The counter is initialised with $ 0 $ and the loop body is instantiated with \textcolor{blue}{$ \TCall{\Expr} $} to call the first loop iteration.
Else, the loop is skipped and we project the loop continuation $ G_2 $ on $ \Role[p] $.

Projection of the remaining operators is given in \cite{petersNestmannWagner23}.
We restrict our attention to projectable and well-formed types, as defined in \cite{bartlLinnePetersTecRep25}.

%%%%%%%%%%%%%%%%%%%%%%
%  Failure Patterns  %
%%%%%%%%%%%%%%%%%%%%%%

\section{A Semantics with Failure Patterns for Global Escape Loops}
\label{sec:failurePatterns}

Before we describe the semantics, we introduce substitution and structural congruence as auxiliary concepts.
The application of a substitution $ \Subst{\Args[y]}{\Args} $ on a term $ A $, denoted as $ A\Subst{\Args[y]}{\Args} $, is defined as the result of replacing all free occurrences of $ \Args $ in $ A $ by $ \Args[y] $, possibly applying alpha-conversion to avoid capture or name clashes. For all names $ n \in \names \setminus \Set{ \Args } $ the substitution behaves as the identity mapping. We use substitution on types as well as processes and naturally extend substitution to the substitution of variables by terms (to unfold recursions) and names by expressions (to instantiate a bound name with a received value).

We use structural congruence to abstract from syntactically different processes with the same meaning, where $ \equiv $ is the least congruence that satisfies alpha conversion and the rules:
\[ \begin{array}{c}
	\PT \mid \PEnd \equiv \PT
	\hspace{2em}
	\PT_1 \mid \PT_2 \equiv \PT_2 \mid \PT_1
	\hspace{2em}
	\PT_1 \mid \left( \PT_2 \mid \PT_3 \right) \equiv \left( \PT_1 \mid \PT_2 \right) \mid \PT_3\
	\hspace{2em}
	\PRep{\ProcV, \textcolor{blue}{\Counter = n}}{\PEnd} \equiv \PEnd\\
	\PRes{\Args}{\PEnd} \equiv \PEnd
	\hspace{2em}
	\PRes{\Args}{\PRes{\Args[y]}{\PT}} \equiv \PRes{\Args[y]}{\PRes{\Args}{\PT}}
	\hspace{2em}
	\PRes{\Args}{\left( \PT_1 \mid \PT_2 \right)} \equiv \PT_1 \mid \PRes{\Args}{\PT_2} \quad \text{if } \Args \notin \FreeNames{\PT_1}
\end{array} \]

\begin{figure}[t]
	\centering
	\renewcommand{\tabcolsep}{2pt}
	\begin{tabular}{ll}
		(\textsf{Init}) & $ \PReq{\Chan[a]}{\Role[n]}{\Chan}{\PT_{\Role[n]}} \mid \prod_{1 \leq \Role[i] \leq \Role[n] - 1} \PAcc{\Chan[a]}{\Role[i]}{\Chan}{\PT_{\Role[i]}} \step \PRes{\Chan}{\left( \prod_{1 \leq \Role[i] \leq \Role[n]} \PT_{\Role[i]} \mid \prod_{1 \leq \Role[i], \Role[j] \leq \Role[n], \Role[i] \neq \Role[j]} \MQ{\Chan}{\Role[i]}{\Role[j]}{\emptyList} \right)} $ \hfill if $ \Chan[a] \neq \Chan $\\
		(\textsf{RSend}) & $ \PSendR{\Chan}{\Role_1}{\Role_2}{\Expr[e]}{\PT} \mid \MQ{\Chan}{\Role_1}{\Role_2}{\Queue} \step \PT \mid \MQ{\Chan}{\Role_1}{\Role_2}{\Queue\#\MessR{\Expr[v]}} $ \hfill if $ \Eval{\Expr[e]} = \Expr[v] $\\
		(\textsf{RGet}) & $ \PGetR{\Chan}{\Role_1}{\Role_2}{\Args}{\PT} \mid \MQ{\Chan}{\Role_2}{\Role_1}{\MessR{\Expr[v]}\#\Queue} \step \PT\Subst{\Expr[v]}{\Args} \mid \MQ{\Chan}{\Role_2}{\Role_1}{\Queue} $\\
		(\textsf{USend}) & $ \PSendU{\Chan}{\Role_1}{\Role_2}{\Label}{\Expr[e]}{\PT} \mid \MQ{\Chan}{\Role_1}{\Role_2}{\Queue} \step \PT \mid \MQ{\Chan}{\Role_1}{\Role_2}{\Queue\#\MessU{\Label}{\Expr[v]}} $ \hfill if $ \Eval{\Expr[e]} = \Expr[v] $\\
		(\textsf{UGet}) & $ \PGetU{\Chan}{\Role_1}{\Role_2}{\Label}{\Expr[dv]}{\Args}{\PT} \mid \MQ{\Chan}{\Role_2}{\Role_1}{\MessU{\Label'}{\Expr[v]}\#\Queue} \step \PT\Subst{\Expr[v]}{\Args} \mid \MQ{\Chan}{\Role_2}{\Role_1}{\Queue} $ \hfill if $ \Label \compL \Label' $, $ \fpUGet{\left( \Chan, \Role_1, \Role_2, \Label', \ldots \right)} $\\
		(\textsf{USkip}) & $ \PGetU{\Chan}{\Role_1}{\Role_2}{\Label}{\Expr[dv]}{\Args}{\PT} \step \PT\Subst{\Expr[dv]}{\Args} $ \hfill if $ \fpUSkip{\left( \Chan, \Role_1, \Role_2, \Label, \ldots \right)} $\\
		(\textsf{ML}) & $ \MQ{\Chan}{\Role_1}{\Role_2}{\MessU{\Label}{\Expr[v]}\#\Queue} \step \MQ{\Chan}{\Role_1}{\Role_2}{\Queue} $ \hfill if $ \fpML{\left( \Chan, \Role_1, \Role_2, \Label, \ldots \right)} $\\
		(\textsf{RSel}) & $ \PSelR{\Chan}{\Role_1}{\Role_2}{\Label}{\PT} \mid \MQ{\Chan}{\Role_1}{\Role_2}{\Queue} \step \PT \mid \MQ{\Chan}{\Role_1}{\Role_2}{\Queue\#\MessBR{\Label}} $\\
		(\textsf{RBran}) & $ \PBranR{\Chan}{\Role_1}{\Role_2}{\Set{ \Label_i.\PT_i }_{i \in \indexSet}} \mid \MQ{\Chan}{\Role_2}{\Role_1}{\MessBR{\Label}\#\Queue} \step \PT_j \mid \MQ{\Chan}{\Role_2}{\Role_1}{\Queue} $ \hfill if $ \Label \compL \Label_j $, $ j \in \indexSet $\\
		(\textsf{WSel}) & $ \PSelW{\Chan}{\Role}{\Role[R]}{\Label}{\PT} \mid \prod_{\Role_i \in \Role[R]} \MQ{\Chan}{\Role}{\Role_i}{\Queue_i} \step \PT \mid \prod_{\Role_i \in \Role[R]} \MQ{\Chan}{\Role}{\Role_i}{\Queue_i\#\MessBW{\Label}} $\\
		(\textsf{WBran}) & $ \PBranW{\Chan}{\Role_1}{\Role_2}{\Set{ \Label_i.\PT_i }_{i \in \indexSet, \LabelD}} \mid \MQ{\Chan}{\Role_2}{\Role_1}{\MessBW{\Label}\#\Queue} \step \PT_j \mid \MQ{\Chan}{\Role_2}{\Role_1}{\Queue} $ \hfill if $ \Label \compL \Label_j $, $ j \in \indexSet $\\
		(\textsf{WSkip}) & $ \PBranW{\Chan}{\Role_1}{\Role_2}{\Set{ \Label_i.\PT_i }_{i \in \indexSet, \LabelD}} \step \PTD $ \hfill if $ \fpWSkip{\left( \Chan, \Role_1, \Role_2, \ldots \right)} $\\
		\textcolor{blue}{(\textsf{LStep})} & \textcolor{blue}{$ \PLoop{\PLoopHeader{\Chan}{\Role}{\Role[R]}{\Expr}{\Counter}{n}}{\Args}{\PT_0}{\PT_1}{\Args[y]}{\PT_2} \mid \PT[Q] \step \PLoop{\PLoopHeader{\Chan}{\Role}{\Role[R]}{\Expr}{\Counter}{n}}{\Args}{\PT_0}{\PT_1'}{\Args[y]}{\PT_2} \mid \PT[Q]' $}\\
		& \hfill \textcolor{blue}{if $ \PT_1 \mid \PT[Q] \step \PT_1' \mid \PT[Q]' $, $ \OnlyMQ{\Role[r]}{\Role[R]}{\PT[Q], \PT[Q]'} $}\\
		\textcolor{blue}{(\textsf{LCall})} & \textcolor{blue}{$ \PLoop{\PLoopHeader{\Chan}{\Role}{\Role[R]}{\Expr}{\Counter}{n}}{\Args}{\PT_0}{\PCall{\Expr_l}{\Expr_v}}{\Args[y]}{\PT_2} \step{} $}\\
		& \textcolor{blue}{$ \PLoop{\PLoopHeader{\Chan}{\Role}{\Role[R]}{\Expr}{\Counter}{\Eval{n + 1}}}{\Args}{\PT_0}{\left( \PT_0\Subst{n}{\Counter} \right)\Subst{\Expr[v]}{\Args}}{\Args[y]}{\PT_2} $} \hfill \textcolor{blue}{if $ \Eval{\Expr} = \Eval{\Expr_l} $, $ \Eval{\Expr_v} = \Expr[v] $}\\
		\textcolor{blue}{(\textsf{LExitS})} & \textcolor{blue}{$ \PLoop{\PLoopHeader{\Chan}{\Role}{\Role[R]}{\Expr}{\Counter}{n}}{\Args}{\PT_0}{\Exit{\Expr_l}{\Expr_v}}{\Args[y]}{\PT_2} \mid \prod_{\Role_i \in \Role[R]} \MQ{\Chan}{\Role}{\Role_i}{\Queue_i} \step{} $}\\
		& \textcolor{blue}{$ \PT_2\Subst{\Expr[v]}{\Args[y]} \mid \prod_{\Role_i \in \Role[R]} \MQ{\Chan}{\Role}{\Role_i}{\Queue_i \# \Exit{\Expr[id]}{\Expr[v]}} $} \hfill \textcolor{blue}{if $ \Eval{\Expr[e]} = \Eval{\Expr[e_l]} = \Expr[id] $, $ \Eval{\Expr_v} = \Expr[v] $}\\
		\textcolor{blue}{(\textsf{LExitG})} & \textcolor{blue}{$ \PLoop{\PLoopHeader{\Chan}{\Role}{\Role[R]}{\Expr}{\Counter}{n}}{\Args}{\PT_0}{\PT_1}{\Args[y]}{\PT_2} \mid \MQ{\Chan}{\Role'}{\Role}{\Exit{\Expr[id]}{\Expr[v]}\#\Queue} \step{} $}\\
		& \textcolor{blue}{$ \PT_2\Subst{\Expr[v]}{\Args[y]} \mid \MQ{\Chan}{\Role'}{\Role}{\Queue} $} \hfill \textcolor{blue}{if $ \Eval{\Expr} = \Expr[id] $, $ \Role' \in \Role[R] $}\\
		\textcolor{blue}{(\textsf{EDrop})} & \textcolor{blue}{$ \MQ{\Chan}{\Role_2}{\Role_1}{\Exit{\Expr[id]}{\Expr[v]}\#\Queue} \step \MQ{\Chan}{\Role_2}{\Role_1}{\Queue} $} \hfill \textcolor{blue}{if $ \fpDrop{\left( \Role, \Expr[id] \right)} $}\\
		(\textsf{Crash}) & $ \PT \step \PCrash $ \hfill if $ \fpCrash(\PT, \ldots) $\\
		(\textsf{If-T}) & $ \PITE{\Expr}{\PT}{\PT'} \step \PT $ \hfill if $ \Eval{\Expr} $ is true\\
		(\textsf{If-F}) & $ \PITE{\Expr}{\PT}{\PT'} \step \PT' $ \hfill if $ \Eval{\Expr} $ is false\\
		(\textsf{Deleg}) & $ \PDelA{\Chan}{\Role_1}{\Role_2}{\AT{\Chan'}{\Role}}{\PT} \mid \MQ{\Chan}{\Role_1}{\Role_2}{\Queue} \step \PT \mid \MQ{\Chan}{\Role_1}{\Role_2}{\Queue\#\AT{\Chan'}{\Role}} $\\
		(\textsf{SRecv}) & $ \PDelB{\Chan}{\Role_1}{\Role_2}{\AT{\Chan'}{\Role}}{\PT} \mid \MQ{\Chan}{\Role_2}{\Role_1}{\AT{\Chan''}{\Role'}\#\Queue} \step \PT\Subst{\Chan''}{\Chan'}\Subst{\Role'}{\Role} \mid \MQ{\Chan}{\Role_1}{\Role_2}{\Queue} $\\
		(\textsf{Par}) & $ \PT_1 \mid \PT_2 \step \PT_1' \mid \PT_2 $ \hfill if $ \PT_1 \step \PT_1' $\\
		(\textsf{Res}) & $ \PRes{\Args}{\PT} \step \PRes{\Args}{\PT'} $ \hfill if $ \PT \step \PT' $\\
		(\textsf{Rec}) & $ \PRep{\ProcV, \textcolor{blue}{\Counter = n}}{\PT} \step \left( \PT\textcolor{blue}{\Subst{n}{\Counter}} \right) \Subst{\PRep{\ProcV, \textcolor{blue}{\Counter = \Eval{n + 1}}}{\PT}}{\ProcV} $\\
		(\textsf{Struc}) & $ \PT_1 \step \PT_1' $ \hfill if $ \PT_1 \equiv \PT_2 $, $ \PT_2 \step \PT_2' $, $ \PT_2' \equiv \PT_1' $
	\end{tabular}
	\setlength{\belowcaptionskip}{-1ex}
	\caption{Reduction Rules ($ \step $) of Fault-Tolerant Processes with \textcolor{blue}{Global Escape Loops}.}
	\label{fig:semantics}
\end{figure}

For the reduction semantics in Figure~\ref{fig:semantics} we start with the rules of fault-tolerant processes from \cite{petersNestmannWagner23} that we extend with the rules for our new loops (in blue colour).
Similar to \cite{hondaYoshidaCarbone16}, session initialisation is synchronous and communication within a session is asynchronous using message queues.
The rules are standard except for the six failure patterns (five patterns from \cite{petersNestmannWagner23} and one new pattern \textcolor{blue}{$ \fpDrop $} for loops) and three rules for system failures: (\textsf{Crash}) for \emph{crash failures}, (\textsf{ML}) for \emph{message loss}, and the new rule \textcolor{blue}{(\textsf{EDrop})} that allows to drop $ \mathtt{exit} $-messages of loops.
\emph{Failure patterns} are predicates that we deliberately choose not to define here (see below).
They allow us to provide information about the underlying communication medium and the reliability of processes.

Rule~(\textsf{Init}) initialises a session with $ \Role[n] $ roles.
Session initialisation introduces a fresh session channel and unguards the participants of the session.
Finally, the message queues of this session are initialised with the empty list under the restriction of the session channel.

Rule~(\textsf{RSend}) implements an asynchronous \strongR message transmission.
As a result, the value $ \Expr[v] = \Eval{\Expr} $ is wrapped in a message and added to the end of the corresponding message queue and the continuation of the sender is unguarded.
Rule~(\textsf{USend}) is the counterpart of (\textsf{RSend}) for \unrel senders.
(\textsf{RGet}) consumes a message that is marked as \strongR with the index $ \operatorname{r} $ from the head of the respective message queue and replaces in the unguarded continuation of the receiver the bound variable $ \Args $ by the received value $ \Expr[v] $.

There are two rules for the reception of a message in an \unrel communication that are guided by failure patterns.
Rule~(\textsf{UGet}) is similar to Rule~(\textsf{RGet}), but specifies a failure pattern $ \fpUGet $ to decide whether this step is allowed.
This failure pattern could, \eg, be used to reject messages that are too old.
The condition $ \Label \compL \Label' $ ensures that the static information in the transmitted label matches the expectation specified in the label of the receiver to avoid communication mismatches.
The Rule~(\textsf{USkip}) allows to skip the reception of a message in an \unrel communication using a failure pattern $ \fpUSkip $ and instead substitutes the bound variable $ \Args $ in the continuation with the default value $ \Args[dv] $.
The failure pattern $ \fpUSkip $ tells us whether a reception can be skipped (\eg via failure detector).

Rule~(\textsf{RSel}) puts the label $ \Label $ selected by $ \Role_1 $ at the end of the message queue towards $ \Role_2 $.
Its \weakR counterpart (\textsf{WSel}) is similar, but puts the label at the end of all relevant message queues.
With (\textsf{RBran}) a label is consumed from the top of a message queue and the receiver moves to the indicated branch.
There are again two \weakR counterparts of (\textsf{RBran}).
Rule~(\textsf{WBran}) is similar to (\textsf{RBran}), whereas (\textsf{WSkip}) allows $ \Role_1 $ to skip the message and to move to its default branch if the failure pattern $ \fpWSkip $ holds.
The requirement $ \Label \compL \Label_j $ in \textsf{RBran} and \textsf{WBran} ensures as usual that indeed the branch specified by the message at the queue is picked by the receiver.
Note that this branch has to be identified by the statically available information in the respective labels.

With \textcolor{blue}{(\textsf{LStep})} the body of a loop may (1)~send a message to a message queue, (2)~receive a message from a queue, (3)~resolve a conditional, or (4)~skip an outer loop-construct of nested loops to perform an output, input, call another loop iteration, or exit a loop.
Therefore, the predicate \textcolor{blue}{$ \OnlyMQ{\Role}{\Role[R]}{\PT[Q], \PT[Q]'} $} checks that $ \PT[Q] $ and $ \PT[Q]' $ consist only of message queues from $ \Role $ into roles within $ \Role[R] $ or the other way around.
Rule~\textcolor{blue}{(\textsf{LCall})} puts loop $ \Eval{\Expr} $ onto another iteration, where $ \Counter $ is replaced by the current counter value $ n $ and $ \Args $ is instantiated with $ \Eval{\Expr_v} = \Expr[v] $ in the loop program $ {\left( \Args \right)}.\PT_0 $.
Additionally, the counter is increased by 1.
The side condition $ \Eval{\Expr} = \Eval{\Expr_l} $ ensures that the correct loop is iterated.
Role $ \Role $ can terminate its loop $ \Eval{\Expr} = \Eval{\Expr_l} = \Expr[id] $ with \textcolor{blue}{(\textsf{LExitS})}.
This step reduces $ \Role $ to its loop continuation $ {\left( \Args[y] \right)}.\PT_2 $, where $ \Args[y] $ is instantiated with $ \Eval{\Expr_v} = \Expr[v] $.
It then adds the message \textcolor{blue}{$ \Exit{\Expr[id]}{\Expr[v]} $} to the message queues from $ \Role $ to all roles in $ \Role[R] $.
Upon receiving \textcolor{blue}{$ \Exit{\Expr[id]}{\Expr[v]} $} in rule \textcolor{blue}{(\textsf{LExitG})}, role $ \Role $ is induced to also terminate its loop $ \Expr[id] $ and continue with its loop continuation instantiated with $ \Expr[v] $.

The Rules~(\textsf{Crash}) for \emph{crash failures} and (\textsf{ML}) for \emph{message loss}, describe failures of a system.
With Rule~(\textsf{Crash}), $ \PT $ can crash if $ \fpCrash $, where $ \fpCrash $ can \eg model immortal processes or global bounds on the number of crashes.
(\textsf{ML}) allows to drop an \unrel message if the failure pattern $ \fpML $ is valid.
$ \fpML $ allows, \eg, to implement safe channels that never lose messages or a global bound on the number of lost messages.
Rule~\textcolor{blue}{(\textsf{EDrop})}, similarly allows to \emph{drop} a message from a queue, but it does not implement a failure.
Instead it allows us to drop $ \mathtt{exit} $-messages of already terminated loops, \ie $ \fpDrop $ checks whether the loop mentioned by the $ \mathtt{exit} $-message of the considered role is already terminated and only in this case allows to drop the message.
Since a loop $ \Expr[id] $ is run concurrently by several roles of which each role runs its local version of the loop $ \Expr[id] $, it cannot be avoided that several roles may actively terminate their loop concurrently, causing several $ \mathtt{exit} $-messages for the same loop.

The remaining reduction rules for conditionals, delegation, parallel composition, restriction, recursion, and structural congruence are standard, except for the counter in unfolding recursion.

We deliberately do not specify failure patterns, although we usually assume that the failure patterns $ \fpUGet $, $ \fpUSkip $, $ \fpWSkip $, and \textcolor{blue}{$ \fpDrop $} use only local information, whereas $ \fpML $ and $ \fpCrash $ may use global information of the system in the current run.
We provide these predicates to allow for the implementation of system requirements or abstractions like failure detectors that are typical for distributed algorithms.
Directly including them in the semantics has the advantage that all traces satisfy the corresponding requirements, \ie all traces are valid \wrt the assumed system requirements.
An example for the instantiation of these patterns is given implicitly via the Conditions~\ref{cond:all}.\ref{cond:crash}--\ref{cond:all}.\ref{cond:fpDropB} in Section~\ref{sec:typing} and explicitly in Section~\ref{sec:example}.
If we instantiate the patterns $ \fpUGet $ with true, the patterns $ \fpUSkip $, $ \fpWSkip $, $ \fpCrash $, $ \fpML $ with false, and the pattern \textcolor{blue}{$ \fpDrop $} such that it is true whenever the mentioned loop is terminated by the mentioned role, then we obtain a system without failures.
In contrast, the instantiation of \textcolor{blue}{$ \fpDrop $} as above and the other five patterns with true results in a system, where failures can happen completely non-deterministically at any time.

Note that we keep the failure patterns abstract and do not model how to check them in producing runs.
Indeed system requirements such as bounds on the number of processes that can crash usually cannot be checked, but result from observations, \ie system designers ensure that a violation of this bound is very unlikely and algorithm designers are willing to ignore these unlikely events.
In particular, $ \fpML $ and $ \fpCrash $ are thus often implemented as oracles for verification, whereas \eg $ \fpUSkip $ and $ \fpWSkip $ are often implemented by system specific time-outs.
Note that we are talking about implementing these failure patterns and not formalising them.
Failure patterns are abstractions of real world system requirements or software.
We implement them by conditions providing the necessary guarantees that we need in general (\ie for subject reduction and progress) or for the verification of concrete algorithms.
In practice, we expect that the systems on which the verified algorithms are running satisfy the respective conditions.
Accordingly, the session channels, roles, labels, processes, and loop-identifiers mentioned in Figure~\ref{fig:semantics} are not parameters of the failure patterns, but just a vehicle to more formally specify the conditions on failure patterns in Section~\ref{sec:typing}.
An implementation may or may not use these information to implement these patterns but may also use other information such as runtime information about time or the number of processes, as indicated by the \ldots in failure patterns in Figure~\ref{fig:semantics} such as $ \fpCrash(\PT, \ldots) $.

Similarly, \strongR and \weakR interactions in potentially faulty systems are abstractions.
They are usually implemented by handshakes and redundancy; replicated servers against crash failures and retransmission of late messages against message loss.
Algorithm designers have to be aware of the additional costs of these interactions.

The following toy-example illustrates nested loops in types and projection.
A more interesting example with communication is given in Sections~\ref{sec:example}.
\begin{align*}
	\GT &\deff \GRep{\TypeV, \Counter_1}{\GLoop{\Set{\Role[1]}}{\Counter_1}{\Counter_2}{\nat}{
		\GLoop{\Set{\Role[1]}}{\left( \Counter_1, \Counter_2 \right)}{\Counter_3}{\nat}{\GEnd}{\nat}{\TCall{\Counter_1}}
	}{\nat}{\TypeV}}\\
	\Proj{\GT}{\Role[1]} &= \LRep{\TypeV, \Counter_1 = 0}{\LLoop{\emptyset}{\Counter_1}{\Counter_2}{0}{\nat}{
		\LLoop{\emptyset}{\left( \Counter_1, \Counter_2 \right)}{\Counter_3}{0}{\nat}{\LEnd}{\TCall{\left( \Counter_1, \Counter_2 \right)}}{\nat}{\TCall{\Counter_1}}
	}{\TCall{\Counter_1}}{\nat}{\TypeV}}
\end{align*}
To ensure that the loops are uniquely identified in all unfoldings of the surrounding recursion and the outer loop, their identifiers $ \Counter_1 $ and $ \left( \Counter_1, \Counter_2 \right) $ are build from counters.
This type can be implemented as:
\begin{align*}
	\PT \deff {} & \PRep{\ProcV, \Counter_1 = 0}{\PT_{\Counter_1}} \\
	\PT_{\Counter_1} \deff {} & \PLoop{\PLoopHeader{\Chan}{\Role[1]}{\emptyset}{\Counter_1}{\Counter_2}{0}}{\Args}{\PT_{\Counter_1, \Counter_2}(\Args)}{\PCall{\Counter_1}{0}}{\Args[y]}{\ProcV} \\
	\PT_{\Counter_1, \Counter_2}(\Args) \deff {} & \PLoop{\PLoopHeader{\Chan}{\Role[1]}{\emptyset}{\left( \Counter_1, \Counter_2 \right)}{\Counter_3}{0}}{\Args'}{\Exit{\left( \Counter_1, \Counter_2 \right)}{\Args' + 1}}{\PCall{\left( \Counter_1, \Counter_2 \right)}{\Args + 1}}{\Args[y]'}{\PT_{\Counter_1, \mathtt{cont}}(\Args[y]')} \\
	\PT_{\Counter_1, \mathtt{cont}}(\Args[y]') \deff {} & \PITE{\Args[y'] < 5}{\PCall{\Counter_1}{\Args[y]' + 1}}{\Exit{\Counter_1}{\Args[y]' + 1}}\\
	\PT &\step \PT_{0}\Subst{\PRep{\ProcV, \Counter_1 = 1}{\PT_{\Counter_1}}}{\ProcV}\\
	&\step \PLoop{\PLoopHeader{\Chan}{\Role[1]}{\emptyset}{0}{\Counter_2}{1}}{\Args}{\PT_{0, \Counter_2}(\Args)}{\PT_{0, 0}(0)}{\Args[y]}{\PRep{\ProcV, \Counter_1 = 1}{\PT_{\Counter_1}}}\\
	&\steps \PLoop{\PLoopHeader{\Chan}{\Role[1]}{\emptyset}{0}{\Counter_2}{1}}{\Args}{\PT_{0, \Counter_2}(\Args)}{\PCall{0}{2 + 1}}{\Args[y]}{\PRep{\ProcV, \Counter_1 = 1}{\PT_{\Counter_1}}} \\
	&\step \PLoop{\PLoopHeader{\Chan}{\Role[1]}{\emptyset}{0}{\Counter_2}{2}}{\Args}{\PT_{0, \Counter_2}(\Args)}{\PT_{0, 1}(3)}{\Args[y]}{\PRep{\ProcV, \Counter_1 = 1}{\PT_{\Counter_1}}} \\
	&\steps \PLoop{\PLoopHeader{\Chan}{\Role[1]}{\emptyset}{0}{\Counter_2}{2}}{\Args}{\PT_{0, \Counter_2}(\Args)}{\Exit{0}{5 + 1}}{\Args[y]}{\PRep{\ProcV, \Counter_1 = 1}{\PT_{\Counter_1}}} \\
	&\step \PRep{\ProcV, \Counter_1 = 1}{\PT_{\Counter_1}}\Subst{6}{\Args[y]}
	\step \PT_{1}\Subst{\PRep{\ProcV, \Counter_1 = 2}{\PT_{\Counter_1}}}{\ProcV}
\end{align*}

%%%%%%%%%%%%%%%%%%%%%%%%%%%%%%%%%%%%%
%  Typing Fault-Tolerant Processes  %
%%%%%%%%%%%%%%%%%%%%%%%%%%%%%%%%%%%%%

\section{Typing Fault-Tolerant Processes}
\label{sec:typing}

The type of processes is checked using typing rules that define the derivation of type judgments.
Within type judgements, the type information are stored in type environments.

\begin{definition}[Type Environments]
	\label{def:typeEnvironments}
	The \emph{global}, \textcolor{blue}{\emph{loop}} and \emph{session environments} are given by
	\begin{align*}
		\Gamma & \deffTerms
			\emptyset
			\sepTerms \Gamma \compS \Typed{\Args}{\Sort}
			\sepTerms \Gamma \compS \Typed{\Chan[a]}{\GT}
			\sepTerms \Gamma \compS \Typed{\Label}{\Sort}\\
		\textcolor{blue}{\Theta} & \deffTerms
			\emptyset
			\sepTerms \Theta \compS \Typed{\ProcV}{\AT{\Chan}{\Role}\TypeV}
			\sepTerms \textcolor{blue}{\Typed{\Expr}{\AT{\Chan}{\Role}{\left \langle \Sort_0, \Sort_2 \right\rangle}}}\\
		\Delta & \deffTerms
			\emptyset
			\sepTerms \Delta \compS \Typed{\AT{\Chan}{\Role}}{\LT}
			\sepTerms \Delta \compS \MQ{\Chan}{\Role_1}{\Role_2}{\MT}
	\end{align*}
\end{definition}

Global environments with assignments $ \Typed{\Args}{\Sort} $ of values to sorts, $ \Typed{\Chan[a]}{\GT} $ of shared channels $ \Chan[a] $ to global types (for session initialisation), and $ \Typed{\Label}{\Sort} $ of labels to sorts as well as session environments with assignments $ \Typed{\AT{\Chan[s]}{\Role}}{\LT} $ of actors to local types and $ \MQ{\Chan[s]}{\Role_1}{\Role_2}{\MT} $ of message queues to a list of message types are inherited from \cite{petersNestmannWagner23}.
We move assignments $ \Typed{\ProcV}{\AT{\Chan}{\Role}\TypeV} $ of process variables to actors and type variables (to check standard recursion) to the new \textcolor{blue}{loop environments} that also contains assignments $ \textcolor{blue}{\Typed{\Expr}{\AT{\Chan}{\Role}{\left \langle \Sort_0, \Sort_2 \right\rangle}}} $ of loop identifiers to actors and sorts (for the values used to call and exit a loop).
Loop environments are used to list active recursion and loops inside their respective bodies.

We write $ \Args \sharp \Gamma $, $ \Args \sharp \Theta $, and $ \Args \sharp \Delta $ if $ \Args $ does not occur in $ \Gamma $, $ \Theta $, and $ \Delta $, respectively.
We use $ \compS $ to add an assignment provided that the new assignment is not in conflict with the type environment.
More precisely, $ \Gamma \compS \Typed{\Args}{\Sort} $ implies $ \Args \sharp \Gamma $, $ \Gamma \compS \Typed{\Label}{\Sort} $ implies $ \Label \sharp \Gamma $, $ \Theta \compS \Typed{\ProcV}{\AT{\Chan}{\Role}\TypeV} $ implies $ \ProcV, \TypeV \sharp \Theta $, $ \textcolor{blue}{\Theta \compS \Typed{\Expr}{\AT{\Chan}{\Role}{\left \langle \Sort_0, \Sort_2 \right\rangle}}} $ implies $ \Expr \sharp \Theta $, $ \Delta \compS \Typed{\AT{\Chan[s]}{\Role}}{\LT} $ implies $ \left( \nexists \LT' \logdot \Typed{\AT{\Chan[s]}{\Role}}{\LT'} \in \Delta \right) $, and $ \Delta \compS \MQ{\Chan[s]}{\Role_1}{\Role_2}{\MT} $ implies $ \left( \nexists \MT' \logdot \MQ{\Chan[s]}{\Role_1}{\Role_2}{\MT'} \in \Delta \right) $.
We naturally extend this operator towards sets, \ie $ \Gamma \compS \Gamma' $ implies $ \left( \forall A \in \Gamma' \logdot \Gamma \compS A \right) $, $ \Theta \compS \Theta' $ implies $ \left( \forall A \in \Theta' \logdot \Theta \compS A \right) $, and $ \Delta \compS \Delta' $ implies $ \left( \forall A \in \Delta' \logdot \Delta \compS A \right) $.
The conditions described for the operator $ \compS $ for global and session environments are referred to as \emph{linearity}.
Accordingly, we denote type environments that satisfy these properties as \emph{linear} and restrict in the following our attention to linear environments.
We abstract in session environments from assignments towards terminated local types, \ie $ \Delta \compS \Typed{\AT{\Chan}{\Role}}{\LEnd} = \Delta $.

\begin{figure}[t]
	\centering
	\[ \begin{array}{c}
		\left( \textsf{Req} \right) \dfrac{\Typed{\Chan[a]}{\GT} \in \Gamma \quad \Length{\Roles{\GT}} = \Role[n] \quad \Gamma, \textcolor{blue}{\Theta} \vdash \PT \triangleright \Delta \compS \Typed{\AT{\Chan}{\Role[n]}}{\Proj{\GT}{\Role[n]}}}{\Gamma, \textcolor{blue}{\Theta} \vdash \PReq{\Chan[a]}{\Role[n]}{\Chan}{\PT} \triangleright \Delta}
		\hspace{2em}
		\left( \textsf{If} \right) \dfrac{\Gamma \Vdash \Typed{\Expr}{\bool} \quad \Gamma, \textcolor{blue}{\Theta} \vdash \PT \triangleright \Delta \quad \Gamma, \textcolor{blue}{\Theta} \vdash \PT' \triangleright \Delta}{\Gamma, \textcolor{blue}{\Theta} \vdash \PITE{\Expr}{\PT}{\PT'} \triangleright \Delta}
		\vspace{0.25em}\\
		\left( \textsf{Acc} \right) \dfrac{\Typed{\Chan[a]}{\GT} \in \Gamma \quad 0 < \Role < \Length{\Roles{\GT}} \quad \Gamma, \textcolor{blue}{\Theta} \vdash \PT \triangleright \Delta \compS \Typed{\AT{\Chan}{\Role}}{\Proj{\GT}{\Role}}}{\Gamma, \textcolor{blue}{\Theta} \vdash \PAcc{\Chan[a]}{\Role}{\Chan}{\PT} \triangleright \Delta}
		\hspace{2em}
		\left( \textsf{End} \right) \dfrac{\textcolor{blue}{\NoLoop{\Theta}}}{\Gamma, \textcolor{blue}{\Theta} \vdash \PEnd \triangleright \emptyset}
		\vspace{0.25em}\\
		\left( \textsf{RSend} \right) \dfrac{\Gamma \Vdash \Typed{\Expr[y]}{\Sort} \quad \Gamma, \textcolor{blue}{\Theta} \vdash \PT \triangleright \Delta \compS \Typed{\AT{\Chan}{\Role_1}}{\LT}}{\Gamma, \textcolor{blue}{\Theta} \vdash \PSendR{\Chan}{\Role_1}{\Role_2}{\Expr[y]}{\PT} \triangleright \Delta \compS \Typed{\AT{\Chan}{\Role_1}}{\LSendR{\Role_2}{\Sort}{\LT}}}
		\hspace{1em}
		\left( \textsf{Rec} \right) \dfrac{\textcolor{blue}{\Gamma \Vdash \Typed{n}{\nat}} \quad \Gamma \compS \textcolor{blue}{\Typed{\Counter}{\nat}}, \textcolor{blue}{\Theta} \compS \Typed{\ProcV}{\AT{\Chan}{\Role}\TypeV} \vdash \PT \triangleright \Typed{\AT{\Chan[s]}{\Role}}{\LT}}{\Gamma, \textcolor{blue}{\Theta} \vdash \PRep{\ProcV, \textcolor{blue}{\Counter = n}}{\PT} \triangleright \Typed{\AT{\Chan[s]}{\Role}}{\LRep{\TypeV, \textcolor{blue}{\Counter = n}}{\LT}}}
		\vspace{0.25em}\\
		\left( \textsf{RGet} \right) \dfrac{\Args \sharp \left( \Gamma, \Delta, \Chan \right) \quad \Gamma \compS \Typed{\Args}{\Sort}, \textcolor{blue}{\Theta} \vdash \PT \triangleright \Delta \compS \Typed{\AT{\Chan[s]}{\Role_1}}{\LT}}{\Gamma, \textcolor{blue}{\Theta} \vdash \PGetR{\Chan}{\Role_1}{\Role_2}{\Args}{\PT} \triangleright \Delta \compS \Typed{\AT{\Chan}{\Role_1}}{\LGetR{\Role_2}{\Sort}{\LT}}}
		\hspace{2em}
		\left( \textsf{Var} \right) \dfrac{}{\Gamma, \textcolor{blue}{\Theta} \compS \Typed{\ProcV}{\AT{\Chan}{\Role}\TypeV} \vdash \ProcV \triangleright \Typed{\AT{\Chan}{\Role}}{\TypeV}}
		\vspace{0.25em}\\
		\left( \textsf{USend} \right) \dfrac{\Gamma \Vdash \Typed{\Expr[y]}{\Sort} \quad \Label \compL \Label' \quad \Typed{\Label'}{\Sort} \in \Gamma \quad \Gamma, \textcolor{blue}{\Theta} \vdash \PT \triangleright \Delta \compS \Typed{\AT{\Chan}{\Role_1}}{\LT}}{\Gamma, \textcolor{blue}{\Theta} \vdash \PSendU{\Chan}{\Role_1}{\Role_2}{\Label}{\Expr[y]}{\PT} \triangleright \Delta \compS \Typed{\AT{\Chan}{\Role_1}}{\LSendU{\Role_2}{\Label'}{\Sort}{\LT}}}
		\hspace{2em}
		\left( \textsf{Par} \right) \dfrac{\Gamma, \textcolor{blue}{\Theta} \vdash \PT \triangleright \Delta \quad \Gamma, \textcolor{blue}{\Theta} \vdash \PT' \triangleright \Delta'}{\Gamma, \textcolor{blue}{\Theta} \vdash \PT \mid \PT' \triangleright \Delta \compS \Delta'}
		\vspace{0.25em}\\
		\left( \textsf{UGet} \right) \dfrac{\Args \sharp \left( \Gamma, \Delta, \Chan \right) \quad \Gamma \Vdash \Typed{\Expr[v]}{\Sort} \quad \Label \compL \Label' \quad \Typed{\Label'}{\Sort} \in \Gamma \quad \Gamma \compS \Typed{\Args}{\Sort}, \textcolor{blue}{\Theta} \vdash \PT \triangleright \Delta \compS \Typed{\AT{\Chan[s]}{\Role_1}}{\LT}}{\Gamma, \textcolor{blue}{\Theta} \vdash \PGetU{\Chan}{\Role_1}{\Role_2}{\Label}{\Expr[v]}{\Args}{\PT} \triangleright \Delta \compS \Typed{\AT{\Chan}{\Role_1}}{\LGetU{\Role_2}{\Label'}{\Sort}{\LT}}}
		\hspace{2em}
		\left( \textsf{Crash} \right) \dfrac{\NotStronglyReliable{\Delta}}{\Gamma, \textcolor{blue}{\Theta} \vdash \PCrash \triangleright \Delta}
		\vspace{0.25em}\\
		\left( \textsf{RSel} \right) \dfrac{j \in \indexSet \quad \Label \compL \Label_j \quad \Gamma, \textcolor{blue}{\Theta} \vdash \PT \triangleright \Delta \compS \Typed{\AT{\Chan}{\Role_1}}{\LT_j}}{\Gamma, \textcolor{blue}{\Theta} \vdash \PSelR{\Chan}{\Role_1}{\Role_2}{\Label}{\PT} \triangleright \Delta \compS \Typed{\AT{\Chan}{\Role_1}}{\LSelR{\Role_2}{\Set{ \Label_i.\LT_i }_{i \in \indexSet}}}}
		\hspace{1em}
		\left( \textsf{WSel} \right) \dfrac{j \in \indexSet \quad \Label \compL \Label_j \quad \Gamma, \textcolor{blue}{\Theta} \vdash \PT \triangleright \Delta \compS \Typed{\AT{\Chan}{\Role}}{\LT_j}}{\Gamma, \textcolor{blue}{\Theta} \vdash \PSelW{\Chan}{\Role}{\Role[R]}{\Label}{\PT} \triangleright \Delta \compS \Typed{\AT{\Chan}{\Role}}{\LSelW{\Role[R]}{\Set{ \Label_i.\LT_i }_{i \in \indexSet}}}}
		\vspace{0.25em}\\
		\left( \textsf{RBran} \right) \dfrac{\forall j \in \indexSet_2\logdot \exists i \in \indexSet_1\logdot \Label_i \compL \Label_j' \wedge \Gamma, \textcolor{blue}{\Theta} \vdash \PT_i \triangleright \Delta \compS \Typed{\AT{\Chan}{\Role_1}}{\LT_j}}{\Gamma, \textcolor{blue}{\Theta} \vdash \PBranR{\Chan}{\Role_1}{\Role_2}{\Set{ \Label_i.\PT_i }_{i \in \indexSet_1}} \triangleright \Delta \compS \Typed{\AT{\Chan}{\Role_1}}{\LBranR{\Role_2}{\Set{ \Label_i'.\LT_i }_{i \in \indexSet_2}}}}
		\hspace{2em}
		\left( \textsf{Res1} \right) \dfrac{\Args \sharp \left( \Gamma, \Delta \right) \quad \Gamma \compS \Typed{\Args}{\Sort}, \textcolor{blue}{\Theta} \vdash \PT \triangleright \Delta}{\Gamma, \textcolor{blue}{\Theta} \vdash \PRes{\Args}{\PT} \triangleright \Delta}
		\vspace{0.25em}\\
		\left( \textsf{WBran} \right) \dfrac{\LabelD \compL \LabelD' \quad \forall j \in \indexSet_2\logdot \exists i \in \indexSet_1\logdot \Label_i \compL \Label_j' \wedge \Gamma, \textcolor{blue}{\Theta} \vdash \PT_i \triangleright \Delta \compS \Typed{\AT{\Chan}{\Role_1}}{\LT_j}}{\Gamma, \textcolor{blue}{\Theta} \vdash \PBranW{\Chan}{\Role_1}{\Role_2}{\Set{ \Label_i.\PT_i }_{i \in \indexSet_1, \LabelD}} \triangleright \Delta \compS \Typed{\AT{\Chan}{\Role_1}}{\LBranW{\Role_2}{\Set{ \Label_i'.\LT_i }_{i \in \indexSet_2, \LabelD'}}}}
		\vspace{0.25em}\\
		\left( \textsf{Deleg} \right) \dfrac{\Gamma, \textcolor{blue}{\Theta} \vdash \PT \triangleright \Delta \compS \Typed{\AT{\Chan}{\Role_1}}{\LT}}{\begin{array}{c} \Gamma, \textcolor{blue}{\Theta} \vdash \PDelA{\Chan}{\Role_1}{\Role_2}{\AT{\Chan'}{\Role}}{\PT} \triangleright {}\\ \Delta \compS \Typed{\AT{\Chan}{\Role_1}}{\LDelA{\Role_2}{\Chan'}{\Role}{\LT'}{\LT}} \compS \Typed{\AT{\Chan'}{\Role}}{\LT'} \end{array}}
		\hspace{2em}
		\left( \textsf{SRecv} \right) \dfrac{\Gamma, \textcolor{blue}{\Theta} \vdash \PT \triangleright \Delta \compS \Typed{\AT{\Chan}{\Role_1}}{\LT} \compS \Typed{\AT{\Chan'}{\Role}}{\LT'}}{\begin{array}{c} \Gamma, \textcolor{blue}{\Theta} \vdash \PDelB{\Chan}{\Role_1}{\Role_2}{\AT{\Chan'}{\Role}}{\PT} \triangleright {}\\ \Delta \compS \Typed{\AT{\Chan}{\Role_1}}{\LDelB{\Role_2}{\Chan'}{\Role}{\LT'}{\LT}} \end{array}}
		\vspace{0.25em}\\
		\textcolor{blue}{\left( \textsf{Loop} \right) \dfrac{\begin{array}{c} \Args, \Args[y] \sharp \left( \Gamma, \Delta, \Chan \right) \quad \Unreliable{\LT_0} \quad \Unreliable{\LT_1} \quad \Gamma \compS \Typed{\Args}{\Sort_0} \compS \Typed{\Counter}{\nat}, \Typed{\Expr}{\AT{\Chan}{\Role}{\left\langle \Sort_0, \Sort_2 \right\rangle}} \vdash \PT_0 \triangleright \Typed{\AT{\Chan}{\Role}}{\LT_0}\\ \Gamma \Vdash \Typed{n}{\nat} \quad \Gamma \compS \Typed{\Counter}{\nat}, \Typed{\Expr}{\AT{\Chan}{\Role}{\left\langle \Sort_0, \Sort_2 \right\rangle}} \vdash \PT_1 \triangleright \Typed{\AT{\Chan}{\Role}}{\LT_1} \quad \Gamma \compS \Typed{\Args[y]}{\Sort_2}, \Theta \vdash \PT_2 \triangleright \Delta \compS \Typed{\AT{\Chan}{\Role}}{\LT_2} \end{array}}{\Gamma, \Theta \vdash \PLoop{\PLoopHeader{\Chan}{\Role}{\Role[R]}{\Expr}{\Counter}{n}}{\Args}{\PT_0}{\PT_1}{\Args[y]}{\PT_2} \triangleright \Delta \compS \Typed{\AT{\Chan}{\Role}}{\LLoop{\Role[R]}{\Expr}{\Counter}{n}{\Sort_0}{\LT_0}{\LT_1}{\Sort_2}{\LT_2}}}}
		\vspace{0.25em}\\
		\textcolor{blue}{\left( \textsf{Call} \right) \dfrac{\Gamma \Vdash \Typed{\Expr_v}{\Sort_0}}{\Gamma, \Theta \compS \Typed{\Expr}{\AT{\Chan}{\Role}{\left\langle \Sort_0, \Sort_2 \right\rangle}} \vdash \PCall{\Expr}{\Expr_v} \triangleright \Typed{\AT{\Chan}{\Role}}{\TCall{\Expr}}}}
		\hspace{0.25em}
		\textcolor{blue}{\left( \textsf{Exit} \right) \dfrac{\Gamma \Vdash \Typed{\Expr_v}{\Sort_2}}{\Gamma, \Theta \compS \Typed{\Expr}{\AT{\Chan}{\Role}{\left\langle \Sort_0, \Sort_2 \right\rangle}} \vdash \Exit{\Expr}{\Expr_v} \triangleright \Typed{\AT{\Chan}{\Role}}{T_1}}}
	\end{array} \]
	\setlength{\abovecaptionskip}{-0.5ex}
	\setlength{\belowcaptionskip}{-1ex}
	\caption{Typing Rules for Fault-Tolerant Systems with \textcolor{blue}{Global Escape Loops}.}
	\label{fig:typingRules}
\end{figure}

A \emph{type judgement} is of the form $ \Gamma, \textcolor{blue}{\Theta} \vdash \PT \triangleright \Delta $, where $ \Gamma $ is a global environment, \textcolor{blue}{$ \Theta $} is a loop environment, $ \PT \in \processes $ is a process, and $ \Delta $ is a session environment.
A process $ \PT $ is \emph{well-typed} \wrt $ \Gamma $ and $ \Delta $ if $ \Gamma \vdash \PT \triangleright \Delta $ can be derived from the rules in the Figures~\ref{fig:typingRules} and \ref{fig:runtimeTypingRules}.
We write $ \NotStronglyReliable{\Delta} $ (or $ \Unreliable{\Delta} $) if for all types $ \LT $ in $ \Delta $ we have $ \NotStronglyReliable{\LT} $ (or $ \Unreliable{\LT} $) and if $ \Delta $ does not contain message queues.
With $ \Gamma \Vdash \Typed{\Expr[y]}{\Sort} $ we check that $ \Expr[y] $ is an expression of the sort $ \Sort $ if all names $ \Args $ in $ \Expr[y] $ are replaced by arbitrary values of sort $ \Sort_{\Args} $ for $ \Typed{\Args}{\Sort_{\Args}} \in \Gamma $.

For the rules in Figure~\ref{fig:typingRules} we adapted the rules of \cite{petersNestmannWagner23} and extended them by rules for loops.
We added the loop environment to all rules that is only relevant for typing recursion and loops.
In (\textsf{End}) we add the condition \textcolor{blue}{$ \NoLoop{\Theta} $} that checks that $ \Theta $ does not contain loop identifiers, to ensure that no branch of a loop program or loop body terminates with $ \PEnd $.

\textcolor{blue}{(\textsf{Loop})} requires the types of a loop program $ T_0 $ and a loop body $ T_1 $ to be unreliable (\textcolor{blue}{$ \Unreliable{T_0} $} and \textcolor{blue}{$ \Unreliable{T_1} $}).
It checks the loop program $ P_0 $ and the loop body $ P_1 $ against their types, but reduces in this check the loop environment to the information for the current loop.
This ensures that $ P_0 $ and $ P_1 $ do not contain free process variables and no calls or exists of surrounding loops.
We do not forbid complete recursions or nested loops inside a loop program/body, where the type system ensures their completion before the end of the loop program/body.
As in recursion via (\textsf{Rec}), we also reduce the session environment to the actor that initiates this loop.
Finally, \textcolor{blue}{(\textsf{Loop})} checks the loop continuation $ P_2 $ against its type $ T_2 $, where $ \Theta $ and $ \Delta $ are not reduced.
Note that to apply this rule, the expression $ \Expr $ used to create the identifier of the loop in the process and the type have to match exactly, \ie are not evaluated.

\looseness=-1
\textcolor{blue}{(\textsf{Call})} is similar to (\textsf{Var}) and checks that the considered recursion or loop is considered active by the loop environment.
Additionally it verifies the sort of the transmitted value.
Also \textcolor{blue}{(\textsf{Exit})} checks the sort of the transmitted value, requires that the current session environment contains only the actor that invoked the considered loop, and that this loop is considered active by the loop environment.
Since \textcolor{blue}{(\textsf{Exit})} does not implement any requirement on the type $ T_1 $, it does intuitively allow to ignore whatever is left of the loop body.

Figure~\ref{fig:runtimeTypingRules} presents the runtime typing rules, \ie the typing rules for processes that may result from steps of a system that implements a global type.
Since it covers only operators that are not part of initial systems, a type checking tool might ignore them.
We need these rules however for the proofs of progress and subject reduction.
Under the assumption that initial systems cannot contain crashed processes, Rule~(\textsf{Crash}) may be moved to the set of runtime typing rules.

\begin{figure}[t]
	\[ \begin{array}{c}
		\left( \textsf{Res2} \right) \dfrac{\begin{array}{c} \Set{ \Typed{\AT{\Chan}{\Role}}{\Proj{\GT}{\Role}} \mid \Role \in \Roles{\GT} } \compS \Set{ \MQ{\Chan}{\Role}{\Role'}{\emptyList} \mid \Role, \Role' \in \Roles{\GT'} \wedge \Role \neq \Role' } \stackrel{\Chan}{\Mapsto} \Delta'\\ \Chan \sharp \left(\Gamma, \Delta \right) \quad \Typed{\Chan[a]}{G} \in \Gamma, \textcolor{blue}{\Theta} \quad \Gamma \vdash \PT \triangleright \Delta \compS \Delta' \end{array}}{\Gamma \vdash \PRes{\Chan[s]}{\PT} \triangleright \Delta}
		\vspace{0.25em}\\
		\left( \textsf{MQComR} \right) \dfrac{\Gamma \Vdash \Typed{\Expr[v]}{\Sort} \quad \Gamma, \textcolor{blue}{\Theta} \vdash \MQ{\Chan}{\Role_1}{\Role_2}{\Queue} \triangleright \MQ{\Chan}{\Role_1}{\Role_2}{\MT}}{\Gamma, \textcolor{blue}{\Theta} \vdash \MQ{\Chan}{\Role_1}{\Role_2}{\MessR{\Expr[v]}\#\Queue} \triangleright \MQ{\Chan}{\Role_1}{\Role_2}{\MessR{\Sort}\#\MT}}
		\vspace{0.25em}\\
		\left( \textsf{MQComU} \right) \dfrac{\Gamma \Vdash \Typed{\Expr[v]}{\Sort} \quad \Label \compL \Label' \quad \Typed{\Label'}{\Sort} \in \Gamma \quad \Gamma, \textcolor{blue}{\Theta} \vdash \MQ{\Chan}{\Role_1}{\Role_2}{\Queue} \triangleright \MQ{\Chan}{\Role_1}{\Role_2}{\MT}}{\Gamma, \textcolor{blue}{\Theta} \vdash \MQ{\Chan}{\Role_1}{\Role_2}{\MessU{\Label}{\Expr[v]}\#\Queue} \triangleright \MQ{\Chan}{\Role_1}{\Role_2}{\MessU{\Label'}{\Sort}\#\MT}}
		\vspace{0.25em}\\
		\left( \textsf{MQBranR} \right) \dfrac{\Label \compL \Label' \quad \Gamma, \textcolor{blue}{\Theta} \vdash \MQ{\Chan}{\Role_1}{\Role_2}{\Queue} \triangleright \MQ{\Chan}{\Role_1}{\Role_2}{\MT}}{\Gamma, \textcolor{blue}{\Theta} \vdash \MQ{\Chan}{\Role_1}{\Role_2}{\MessBR{\Label}\#\Queue} \triangleright \MQ{\Chan}{\Role_1}{\Role_2}{\MessBR{\Label'}\#\MT}}
		\hspace{1em}
		\left( \textsf{MQBranW} \right) \dfrac{\Label \compL \Label' \quad \Gamma, \textcolor{blue}{\Theta} \vdash \MQ{\Chan}{\Role_1}{\Role_2}{\Queue} \triangleright \MQ{\Chan}{\Role_1}{\Role_2}{\MT}}{\Gamma, \textcolor{blue}{\Theta} \vdash \MQ{\Chan}{\Role_1}{\Role_2}{\MessBW{\Label}\#\Queue} \triangleright \MQ{\Chan}{\Role_1}{\Role_2}{\MessBW{\Label'}\#\MT}}
		\vspace{0.25em}\\
		\left( \textsf{MQDeleg} \right) \dfrac{\Gamma, \textcolor{blue}{\Theta} \vdash \MQ{\Chan}{\Role_1}{\Role_2}{\Queue} \triangleright \MQ{\Chan}{\Role_1}{\Role_2}{\MT}}{\Gamma, \textcolor{blue}{\Theta} \vdash \MQ{\Chan}{\Role_1}{\Role_2}{\AT{\Chan'}{\Role}\#\Queue} \triangleright \MQ{\Chan}{\Role_1}{\Role_2}{\AT{\Chan'}{\Role}\#\MT}}
		\hspace{2em}
		\left( \textsf{MQNil} \right) \dfrac{}{\Gamma, \textcolor{blue}{\Theta} \vdash \MQ{\Chan}{\Role_1}{\Role_2}{\emptyList} \triangleright \MQ{\Chan}{\Role_1}{\Role_2}{\emptyList}}
		\vspace{0.25em}\\
		\textcolor{blue}{\left( \textsf{MQExit} \right) \dfrac{\Eval{\Expr} = \Expr[id] \quad \Gamma \Vdash \Typed{\Expr[v]}{\Sort} \quad \Gamma, \Theta \vdash \MQ{\Chan}{\Role_1}{\Role_2}{\Queue} \triangleright \MQ{\Chan}{\Role_1}{\Role_2}{\MT}}{\Gamma, \Theta \vdash \MQ{\Chan}{\Role_1}{\Role_2}{\Exit{\Expr[id]}{\Expr[v]}}\#\Queue \triangleright \MQ{\Chan}{\Role_1}{\Role_2}{\Exit{\Expr}{\Sort}}\#\MT}}
	\end{array} \]
	\setlength{\abovecaptionskip}{-0.5ex}
	\setlength{\belowcaptionskip}{-1ex}
	\caption{Runtime Typing Rules for Fault-Tolerant Systems.}
	\label{fig:runtimeTypingRules}
\end{figure}

Rule (\textsf{Res2}) types sessions that are already initialised and that may have performed already some of the steps described by their global type.
The relation $ \stackrel{\Chan}{\mapsto} $ is given in Figure~5 in \cite{bartlLinnePetersTecRep25} and describes how a session environment evolves alongside reductions of the system, \ie it emulates the reduction steps of processes.
As an example consider the rule
$ \Delta \compS \Typed{\AT{\Chan}{\Role_1}}{\LSendR{\Role_2}{\Sort}{\LT}} \compS \MQ{\Chan}{\Role_1}{\Role_2}{\MT} \stackrel{\Chan}{\mapsto} \Delta \compS \Typed{\AT{\Chan}{\Role_1}}{\LT} \compS \MQ{\Chan}{\Role_1}{\Role_2}{\MT\#\MessR{\Sort}} $
that emulates (\textsf{RSend}).
Let $ \stackrel{\Chan}{\Mapsto} $ denote the reflexive and transitive closure of $ \stackrel{\Chan}{\mapsto} $.

(\textsf{Res2}) and the remaining rules of Figure~\ref{fig:runtimeTypingRules} except for \textcolor{blue}{(\textsf{MQExit})} are from \cite{petersNestmannWagner23} extended by the loop environment \textcolor{blue}{$ \Theta $}.
\textcolor{blue}{(\textsf{MQExit})} checks $ \mathtt{exit} $-messages on a message queue.

We have to prove that our extended type system satisfies the standard properties of \MPST, \ie subject reduction and progress.
Because of the failure patterns in the reduction semantics in Figure~\ref{fig:semantics}, subject reduction and progress do not hold in general.
Instead we have to fix conditions on failure patterns that ensure these properties.
Subject reduction needs one condition on crashed processes and progress requires that no part of the system is blocked.
In fact, different instantiations of these failure patterns may allow for progress.
As in \cite{petersNestmannWagner22, petersNestmannWagner23}, we leave it for future work to determine what kind of conditions on failure patterns or requirements on their interactions are necessary.
Here, we extend the conditions given in \cite{petersNestmannWagner23} by a condition for \textcolor{blue}{$ \fpDrop $}.

\begin{condition}[Failure Pattern]
	\label{cond:all}
	\hfill
	\begin{compactenum}
		\item If $ \fpCrash(\PT, \ldots) $ then $ \NotStronglyReliable{\PT} $. \label{cond:crash}
		\item The failure pattern $ \fpUGet(\Chan, \Role_1, \Role_2, \Label, \ldots) $ is always valid. \label{cond:ugetValid}
		\item The pattern $ \fpML(\Chan, \Role_1, \Role_2, \Label, \ldots) $ is valid iff $ \fpUSkip(\Chan, \Role_2, \Role_1, \Label, \ldots) $ is valid. \label{cond:fpMLifffpUSkip}
		\item If $ \fpCrash(\PT, \ldots) $ and $ \AT{\Chan}{\Role} \in \Actors{\PT} $ is an actor then eventually the pattern $ \fpUSkip(\Chan, \Role_2, \Role, \Label, \ldots) $ and $ \fpWSkip(\Chan, \Role_2, \Role, \Label, \ldots) $ hold for all $ \Role_2, \Label $. \label{cond:fpCrashImpliesSkip}
		\item If $ \fpCrash(\PT, \ldots) $ and $ \AT{\Chan}{\Role} \in \Actors{\PT} $ then eventually $ \fpML(\Chan, \Role_1, \Role, \Label, \ldots) $ for all $ \Role_1, \Label $ and \textcolor{blue}{$ \fpDrop(\Role, \Expr[id]) $}. \label{cond:fpCrashImpliesML}
		\item If $ \fpWSkip(\Chan, \Role_1, \Role_2, \ldots) $ then $ \AT{\Chan}{\Role_2} $ is crashed, \ie the system does no longer contain an actor $ \AT{\Chan}{\Role_2} $ and the message queue $ \MQS{\Chan}{\Role_2}{\Role_1} $ is empty. \label{cond:fpWskip}
		\item \textcolor{blue}{If $ \AT{\Chan}{\Role} $ terminated the loop $ \Expr[id] $ then eventually $ \fpML(\Chan, \Role_1, \Role, \Label, \ldots) $ for all $ \Role_1, \Label $ and $ \fpDrop{\left( \Role, \Expr[id] \right)} $.} \label{cond:fpDropA}
		\item \textcolor{blue}{If $ \fpDrop{\left( \Role, \Expr[id] \right)} $ then $ \Role $ terminated the loop $ \Expr[id] $.} \label{cond:fpDropB}
	\end{compactenum}
\end{condition}

The crash of a process should not block \strongR actions, \ie only processes with $ \NotStronglyReliable{\PT} $ can crash (Condition~\ref{cond:all}.\ref{cond:crash}).
Condition~\ref{cond:all}.\ref{cond:ugetValid} requires that no process can refuse to consume a message on its queue to prevent deadlocks that may arise from refusing a message that is never dropped.
Condition~\ref{cond:all}.\ref{cond:fpMLifffpUSkip} requires that if a message can be dropped from a message queue then the corresponding receiver has to be able to skip this message and vice versa.
Similarly, processes that wait for messages from a crashed process have to be able to skip (Condition~\ref{cond:all}.\ref{cond:fpCrashImpliesSkip}) and all messages of a queue towards a crashed receiver can be dropped (Condition~\ref{cond:all}.\ref{cond:fpCrashImpliesML}).
A \weakR branching request should not be lost.
To ensure that the receiver of such a branching request can proceed if the sender is crashed but is not allowed to skip the reception of the branching request before the sender crashed, we require that $ \fpWSkip(\Chan, \Role_1, \Role_2, \ldots) $ is false as long as $ \AT{\Chan}{\Role_2} $ is alive or messages on the respective queue are still in transit (Condition~\ref{cond:all}.\ref{cond:fpWskip}).
The Conditions~\ref{cond:all}.\ref{cond:fpDropA} and \ref{cond:all}.\ref{cond:fpDropB} ensure that $ \mathtt{exit} $-messages can be dropped after the corresponding loop was terminated but not before.
Moreover, Condition~\ref{cond:all}.\ref{cond:fpDropA} allows to drop messages towards actors of a terminated loop body.
Note that such an actor may also be used in the continuation after the loop.
By adding $ \Expr[id] $ to unreliable messages sent from the loop $ \Expr[id] $, we could more precisely allow to drop only messages that are intended for the loop.
However, the above conditions are sufficient.

It is important to remember that these conditions are minimal assumptions on the system requirements and that system requirements are abstractions.
Parts of them may be realised by actual software-code (which then allows to check them), whereas other parts of the system requirements may not be realised at all but rather observed (which then does not allow to verify them).
Because of that, it is an established method to verify the correctness of algorithms \wrt given system requirements (\eg in \cite{ChandraToueg96,lamport01,Tanenbaum17}), even if these system requirements are not verified and often do not hold in all (but only nearly all) cases.

\emph{Subject reduction} tells us that derivatives of well-typed systems are again well-typed.
This ensures that our formalism can be used to analyse processes by static type checking.
For subject reduction we consider only types that were generated from a set of global types, one for each session, using coherence.
\emph{Coherence} intuitively describes that a session environment captures all local endpoints of a collection of global types.
Since we capture all relevant global types in the global environment, we define
coherence on pairs of global and session environments.

\begin{definition}[Coherence]
	\label{def:coherence}
	The type environments $ \Gamma, \Delta $ are \emph{coherent} if, for all session channels $ \Chan $ in $ \Delta $, there exists a global type $ G $ in $ \Gamma $ such that the restriction of $ \Delta $ on assignments with $ \Chan $ is the set $ \Delta' $ such that:
	\begin{align*}
		\Set{ \Typed{\AT{\Chan}{\Role}}{\Proj{\GT}{\Role}} \mid \Role \in \Roles{\GT} } \compS \Set{ \MQ{\Chan}{\Role}{\Role'}{\emptyList} \mid \Role, \Role' \in \Roles{\GT} } \stackrel{\Chan}{\Mapsto} \Delta'
	\end{align*}
\end{definition}

We use $ \stackrel{\Chan}{\Mapsto} $ in the above definition to define coherence for systems that already performed some steps.

\begin{theorem}[Subject Reduction]
	\label{thm:subjectReduction}
	If $ \Gamma, \Delta $ are coherent, $ \Gamma, \textcolor{blue}{\Theta} \vdash \PT \triangleright \Delta $, and $ \PT \step \PT' $, then there is some $ \Delta' $ such that $ \Gamma, \textcolor{blue}{\Theta} \vdash \PT' \triangleright \Delta' $.
\end{theorem}

The proof is by induction on the derivation of $ \PT \step \PT' $.
In every case, we use the information about the structure of the processes to generate partial proof trees for the respective typing judgement.
Additionally, we use Condition~\ref{cond:all}.\ref{cond:crash} to ensure that the type environment of a crashed process cannot contain the types of reliable communication prefixes.

\emph{Progress} states that no part of a well-typed and coherent system can block other parts, that eventually all matching communication partners are unguarded, that interactions specified by the global type can happen, and that there are no communication mismatches.
Subject reduction and progress together then imply \emph{session fidelity}, \ie that processes behave as specified in their global types.

To ensure that the interleaving of sessions and session delegation cannot introduce deadlocks, we assume an interaction type system as introduced in \cite{BettiniAtall08,hondaYoshidaCarbone16}.
For this type system it does not matter whether the considered actions are \strongR, \weakR, or \unrel.
More precisely, we can adapt the interaction type system of \cite{BettiniAtall08} in a straightforward way to the above session calculus, where \unrel communication and \weakR branching is treated in exactly the same way as \strongR communication/branching, loops are treated in the same way as standard recursion, and exit messages are again ignored, \ie well-typed for arbitrary types.
Remember that loop programs and bodies can act only via the single actor of the loop.
We say that \emph{$ \PT $ is free of cyclic dependencies between sessions} if this interaction type system does not detect any cyclic dependencies.
In this sense fault-tolerance is more flexible than explicit failure handling, which often requires a more substantial revision of the interaction type system to cover the additional dependencies that are introduced \eg by the propagation of faults.

\begin{theorem}[Progress/Session Fidelity]
	\label{thm:progress}
	Let $ \Gamma, \Delta $ be coherent, $ \Gamma, \textcolor{blue}{\Theta} \vdash \PT \triangleright \Delta $, and let $ \PT $ be free of cyclic dependencies between sessions.
	Assume that in the derivation of $ \Gamma, \textcolor{blue}{\Theta} \vdash \PT \triangleright \Delta $, whenever $ \PReq{\Chan[a]}{\Role[n]}{\Chan}{\PT[Q]} $ or $ \PAcc{\Chan[a]}{\Role}{\Chan}{\PT[Q]} $ in $ \PT $, then $ \Typed{\Chan[a]}{\GT} \in \Gamma $, $ \Length{\Roles{G}} = \Role[n] $, and there are $ \PReq{\Chan[a]}{\Role[n]}{\Chan}{\PT[Q]_n} $ as well as $ \PAcc{\Chan[a]}{\Role_i}{\Chan}{\PT[Q]_i} $ in $ \PT $ for all $ 1 \leq \Role_i < \Role[n] $.
	\begin{compactenum}
		\item Then either $ \PT $ does not contain any action prefixes or $ \PT \step \PT' $.
		\item If $ \PT $ does not contain recursion or \textcolor{blue}{loops}, then there exists $ P' $ such that $ \PT \steps \PT' $ and $ \PT' $ does not contain any action prefixes.
	\end{compactenum}
\end{theorem}

The proof of progress relies on the Conditions~\ref{cond:all}.\ref{cond:ugetValid}--\ref{cond:all}.\ref{cond:fpDropB} to ensure that failures cannot block the system: in the failure-free case \unrel messages are eventually received (\ref{cond:all}.\ref{cond:ugetValid}), the receiver of a lost message can skip (\ref{cond:all}.\ref{cond:fpMLifffpUSkip}), no receiver is blocked by a crashed sender (\ref{cond:all}.\ref{cond:fpCrashImpliesSkip}), messages towards receivers that crashed or skipped can be dropped (\ref{cond:all}.\ref{cond:fpCrashImpliesML} + \ref{cond:all}.\ref{cond:fpMLifffpUSkip}), branching requests cannot be ignored (\ref{cond:all}.\ref{cond:fpWskip}), and \textcolor{blue}{$ \mathtt{exit} $-messages can be dropped eventually if and only if the corresponding loop was already terminated (\ref{cond:all}.\ref{cond:fpDropA} + \ref{cond:all}.\ref{cond:fpDropB})}.

%%%%%%%%%%%%%
%  Example  %
%%%%%%%%%%%%%

\section{The Rotating Coordinator Algorithm}
\label{sec:example}

To illustrate the benefits of our global escape loops, we present an implementation of the rotating coordinator algorithm \cite{ChandraToueg96, nestmann07}, which is superior to the version without loops presented in \cite{petersNestmannWagner22, petersNestmannWagner23}.

The rotating coordinator algorithm is a small but not trivial consensus algorithm.
It was designed for systems with crash failures, but the majority of the algorithm can be implemented with \unrel communication.
The goal is that every agent $ \Role[i] $ eventually decides on a proposed belief value, where no two agents decide on different values.
It is a round based algorithm, where each round consists of four phases.
In each round, one process acts as a coordinator decided by round robin, denoted by $ \Role[c] $.
\begin{compactdesc}
	\item[In Phase~1] every agent $ \Role[i] $ sends its current belief to the coordinator $ \Role[c] $.
	\item[In Phase~2] the coordinator waits until it has received at least half of the messages of the current round and then sends the best belief to all other agents.
	\item[In Phase~3] the agents either receive the message of the coordinator or suspect the coordinator to have crashed and reply with ack or nack accordingly. Suspicion can yield false positives.
	\item[In Phase~4] the coordinator waits, as in Phase~2, until it has received at least half of the messages of the current round.
		Then, if at least half of the messages were ack, it sends a \weakR global escape containing the decision.
\end{compactdesc}

It is possible for agents to skip rounds by suspecting the coordinator of the current round and by proceeding to the next round.
There are also no synchronisation fences thus it is possible for the agents to be in different rounds and have messages of different rounds in the system.
Having agents in different rounds makes proving correctness much more difficult.

We use the labels $ \Label[p]_i $ and $ \left( \Label[p]_i, \Counter[r] \right) $, where $ i \in \Set{1, 2, 3} $ specifies the number of the current phase and $ \Counter[r] $ is a natural number that
specifies the current round.  We use $ \Label[p]_i $ as static information and $ \Counter[r] $ as runtime information in the labels.  Therefore, $ \Label[p]_i \compL \left( \Label[p]_i, \Counter[r] \right)
\compL \left( \Label[p]_i, \Counter[r]' \right) $ holds for all $ i $, $ \Counter[r] $, and $ \Counter[r]' $.
The additional runtime information can be used in the failure patterns, \eg to drop outdated messages.
We assume the sorts $ \SortBel = \Set{0,1} $ and $ \SortAck = \Set{\true, \false} $.
Let $ \Role[n] $ be the number of agents.

We start with the specification of the algorithm as a global type.
Let $ {\left( \bigodot_{1 \leq i \leq n} \pi_i \right)}.\GT $ abbreviate $ \pi_1.\ldots.\pi_n.\GT $ to simplify the presentation, where $ \GT $ is a global type and $ \pi_1, \ldots, \pi_n $ are sequences of prefixes.
More precisely, each $ \pi_i $ is of the form $ \pi_{i, 1}.\ldots.\pi_{i, m} $ and each $ \pi_{i, j} $ is a type prefix of the form $ \GComUS{\Role_1}{\Role_2}{\Label}{\Sort} $ or $ \GBranW{\Role}{\Role[R]}{\Label_1.\LT_1 \oplus \ldots \oplus \Label_n.\LT_n \oplus \LabelD} $, where the latter case represents a \weakR branching prefix (as used in \cite{petersNestmannWagner23}) with the branches $ \Label_1, \ldots, \Label_n, \LabelD $, the default branch $ \LabelD $, and where the next global type provides the missing specification for the default case.
\begin{align*}
	\RCG \deff {} & \GLoop{\Set{1, \ldots, \Role[n]}}{1}{\Counter[r]}{\SortBel}{\RCGLoop{\Coordinator}}{\SortBel}{\GEnd}\\
	\RCGLoop{\Role[c]} \deff {}
		& {\Big( \bigodot_{1 \leq \Role[i] \leq \Role[n], \Role[i] \neq \Role[c]} \GComUS{\Role[i]}{\Role[c]}{\Label[p]_1}{\SortBel} \Big)}.
		{\Big( \bigodot_{1 \leq \Role[i] \leq \Role[n], \Role[i] \neq \Role[c]} \GComUS{\Role[c]}{\Role[i]}{\Label[p]_2}{\SortBel} \Big)}.\\
		& {\Big( \bigodot_{1 \leq \Role[i] \leq \Role[n], \Role[i] \neq \Role[c]} \GComUS{\Role[i]}{\Role[c]}{\Label[p]_3}{\SortAck} \Big)}.
		\TCall{1}
\end{align*}
$ \RCG $ specifies a loop with identifier $ 1 $ and counter $ \Counter[r] $, where $ \Counter[r] $ is used as the round number.
The coordinator $ \Role[c] $ of round $ \Counter[r] $ is calculated by $ \Coordinator \deff \left( \Counter[r] \text{ mod } \Counter[n] \right) + 1 $.
Then, $ \RCGLoop{\Role[c]} $ specifies the loop program that implements one round of the algorithm.
The three $ \bigodot $ specify the Phases~1--3 of the algorithm within a single round.
Phase~4 is only specified by $ \TCall{1} $, since there is no $ \mathtt{exit} $ type.

In Phase~1, all processes except the coordinator $ \Role[c] $ transmit a belief to $ \Role[c] $ using label $ \Label[p]_1 $.
In Phase~2, $ \Role[c] $ transmits a belief to all other processes using label $ \Label[p]_2 $.
Then all processes transmit a value of type $ \SortAck $ to the coordinator using label $ \Label[p]_3 $ in Phase~3.
Finally, in Phase~4, the coordinator can terminate the protocol by sending a global escape message containing the decision.
All interactions in the specification are \unrel.

In the following, we implement the algorithm as a process.
Let $ \left( \bigodot_{1 \leq i \leq n} \pi_i \right).\PT $ abbreviate the sequence $ \pi_1.\ldots.\pi_n.\PT $, where $ \PT $ is a process and $ \pi_1, \ldots, \pi_n $ are sequences of prefixes.
\begin{align*}
	\RCSys \deff {} & \PReq{\Chan[a]}{\Role[n]}{\Chan}{\RCP[n]} \mid \prod_{1 \leq \Role[i] < \Role[n]} \PAcc{\Chan[a]}{\Role[i]}{\Chan}{\RCP}\\
	\RCP \deff {} & \PLoop{\PLoopHeader{\Chan}{\Role[i]}{\Set{1, \ldots, \Role[n]} \setminus \Set{\Role[i]}}{1}{\Counter[r]}{0}}{\KnowledgeEntry}{\RCPLoop{r}}{\PCall{1}{\KnowledgeEntry}}{\Args[v]}{\PEnd}\\
	\RCPLoop{r} \deff {} & \PITE{\Role[i] = \Coordinator}{\RCPiC[i]}{\RCPiNC{\Coordinator}}
\end{align*}
$\RCSys$ describes the session initialisation of a system with $\Role[n]$ participants and the (initial) knowledge $ \Knowledge = \Set{\KnowledgeEntry \mid 1 \leq \Role[i] \leq \Role[n]} $, where $ \KnowledgeEntry $ is the initial belief of role $\Role[i]$.
Let $ \Length{\Knowledge} \deff \Length{\Set{\Role[i] \mid \KnowledgeEntry \neq \bot }} $ return the number of non-empty entries.
$ \RCP $ describes a process $ \Role[i] $ in a set of $ \Role[n] $ processes.
Each process is described as a loop with identifier $ 1 $ and counter $ \Counter[r] $, where the loop program executes the round $ \Counter[r] $ of the algorithm.
Once a decision is reached and the loop ends, the loop continuation is instantiated with the decision value.
\begin{align*}
	\RCPiC \deff {} & \Big( \bigodot_{1 \leq \Role[i] \leq \Role[n], \Role[i] \neq \Role[c]} \PGetUS{\Chan}{\Role[c]}{\Role[i]}{\left( \Label[p]_1, \Counter[r] \right)}{\bot}{\KnowledgeEntry} \Big).\\
	& \PITE{\Length{\Knowledge} \geq \ceil{\frac{\Role[n]-1}{2}}}{\RCPiiC{\FuncBest(\Knowledge)}{\FuncBest(\Knowledge)}}{\RCPiiC{\KnowledgeEntry[c]}{\bot}}\\
	\RCPiNC{\Role[c]} \deff {} & \PSendU{\Chan}{\Role[i]}{\Role[c]}{\left( \Label[p]_1, \Counter[r] \right)}{\KnowledgeEntry}{\RCPiiNC}
\end{align*}
In Phase~1, every non-coordinator $ \RCPiNC{\Role[c]} $ sends its own belief via unreliable communication to the coordinator and proceeds to Phase~2.
The coordinator receives (some of) these messages and writes each one into its knowledge vector before proceeding to Phase~2.
If the reception of at least half of the messages was successful, it is updating its belief using the function $ \FuncBest() $ that returns the best belief value.
Otherwise, it continues to use its own belief.
We are using $\big\lceil\frac{\Role[n]-1}{2}\big\rceil$ to check for a majority, since in our implementation processes do not transmit to themselves.
\begin{align*}
	\RCPiiC{\KnowledgeEntry[c]}{\Args} \deff {} & \Big( \bigodot_{1 \leq \Role[i] \leq \Role[n], \Role[i] \neq \Role[c]} \PSendUS{\Chan}{\Role[c]}{\Role[i]}{\left( \Label[p]_2, \Counter[r] \right)}{\Args} \Big).\RCPiiiC\\
	\RCPiiNC \deff {} & \PGetUS{\Chan}{\Role[i]}{\Role[c]}{\left( \Label[p]_2, \Counter[r] \right)}{\bot}{\Args}.\\
	& \PITE{\Args = \bot}{\RCPiiiNC{\KnowledgeEntry}{\false}}{\RCPiiiNC{\Args}{\true}}
\end{align*}
\looseness=-1
In Phase~2, the coordinator sends its updated belief to all other processes via unreliable communication and proceeds.
Note that $ \Args $ is either $ \bot $ or the best belief identified in Phase~1.
If a non-coordinator process successfully receives a belief other than $ \bot $, it updates its own belief with the received value and proceeds to Phase~3, where we use the Boolean value $ \true $ for the acknowledgement.
If the coordinator is suspected to have crashed or $ \bot $ was received, the process proceeds to Phase~3 with the Boolean value $ \false $, signalling nack.
\begin{align*}
	\RCPiiiC \deff {} & \Big( \bigodot_{1 \leq \Role[i] \leq \Role[n], \Role[i] \neq \Role[c]} \PGetUS{\Chan}{\Role[c]}{\Role[i]}{\left( \Label[p]_3, \Counter[r] \right)}{\bot}{\KnowledgeEntry} \Big).\RCPivC\\
	\RCPiiiNC{\KnowledgeEntry}{\Args[b]} \deff {} & \PSendUS{\Chan}{\Role[i]}{\Role[c]}{\left( \Label[p]_3, \Counter[r] \right)}{\Args[b]}.\RCPivNC
\end{align*}
In Phase~3, every non-coordinator sends either ack or nack to the coordinator.
If the coordinator successfully receives the message, it writes the Boolean value at the index of the sender into its knowledge vector.
In case of failure, $\bot$ is used as default.
After that, the processes continue with Phase 4.
\begin{align*}
	\RCPivC \deff {} & \PITE{\FuncCountAck(\Knowledge) \geq \ceil{\frac{\Role[n]-1}{2}}}{\Exit{1}{\KnowledgeEntry[c]}}{\PCall{1}{\KnowledgeEntry[c]}}\\
	\RCPivNC \deff {} & \PCall{1}{\KnowledgeEntry}
\end{align*}
In Phase~4, all non-coordinators move on to the next round.
The coordinator checks if at least half of the non-coordinator roles signalled acknowledgement, utilising the function $ \FuncCountAck() $ to count.
If it received enough acknowledgments, it sends a global escape message containing the decision value, which causes all participants to eventually terminate.
Otherwise, the coordinator continues with the next round.

The main difference between this implementation and the previous version without loops \cite{petersNestmannWagner23} lies in Phase~4.
In the previous version, the coordinator transmitted the decision via broadcasting one of the labels $ \Label[Zero] $, $ \Label[One] $, or $ \LabelD $.
The first two labels represented a decision and terminated the protocol, whereas the default label $ \LabelD $ specified the need for another round:
\begin{align*}
	\RCPivC \deff {} & \PITE{\FuncCountAck(\Knowledge) \geq \ceil{\frac{\Role[n]-1}{2}}}{(\myif \; \Args[v]_{\Role[c]} = 0 \; \mythen \; \PSelW{\Chan}{\Role[c]}{\mathcal{I}}{\Label[Zero]}{\PEnd}\\
	& \myelse \; \PSelW{\Chan}{\Role[c]}{\mathcal{I}}{\Label[One]}{\PEnd})}{\PSelWS{\Chan}{\Role[c]}{\mathcal{I}}{\LabelD}}\\
	\RCPivNC \deff {} & \PBranW{\Chan}{\Role[i]}{\Role[c]}{\Label[Zero].\PEnd \oplus \Label[One].\PEnd \oplus \LabelD}
\end{align*}
where $ \mathcal{I} = \Set{1, \ldots, \Role[n]} \setminus \Set{\Role[c]} $ and the missing continuation after $ \LabelD $ is implemented by the next round.
This caused all non-coordinators to wait for the coordinator's decision before proceeding to the next round.

In our new implementation, presented above, non-coordinators can proceed to the next round immediately after Phase~3.
They can also skip entire rounds by suspecting the coordinator.
Thus, processes can diverge as freely in their rounds as in the original rotating coordinator algorithm \cite{ChandraToueg96}.
Exiting the loop mimics the so-called \emph{reliable broadcast} of the original algorithm.

Chandra and Toueg \cite{ChandraToueg96} introduce the failure detector \FDS that is called \emph{eventually strong}, meaning that
(1) eventually every process that crashes is permanently suspected by every correct process and
(2) there is a time after which some correct process is never suspected by any other process.
We observe that the suspicion of senders is only possible in Phase~3, where processes may suspect the coordinator of the round.
Accordingly, the failure pattern $ \fpUSkip $ implements this failure detector to allow processes to suspect unreliable coordinators in Phase~2, \ie with label~$ \Label[p]_2 $.
In Phase~1 and Phase~3 $ \fpUSkip $ may allow to suspect processes that are not crashed after the coordinator received enough messages.
In all other cases, this pattern eventually returns true iff the respective sender is crashed.
Moreover, $ \fpUSkip $ is true for outdated messages, \ie messages with a round number smaller than the current round of the process.

$ \fpUGet $ returns true.
To prevent the system from becoming blocked, $ \fpML $ and \textcolor{blue}{$ \fpDrop $} eventually return true for messages that cannot be consumed, \ie for messages with label~$ \Label[p]_2 $ that were suspected using \FDS, skipped $ \Label[p]_1 $/$ \Label[p]_3 $-messages, messages from old rounds, and messages after the termination of the loop.
Otherwise, $ \fpML $ and \textcolor{blue}{$ \fpDrop $} returns false.
By the system requirements in \cite{ChandraToueg96}, no messages get lost, but it is realistic to assume that receivers can drop messages of skipped receptions on their incoming message queues.
As there are at least half of the processes required to be correct for this algorithm, we implement $ \fpCrash $ by false if only half of the processes are alive and true otherwise.
These failure patterns satisfy the Conditions~\ref{cond:all}.\ref{cond:crash}--\ref{cond:all}.\ref{cond:fpDropB}.

The proof of termination, agreement, and validity of the algorithm is discussed in \cite{petersNestmannWagner23}.
The main difference is that there may be multiple $ \mathtt{exit} $-messages, but the requirement on the majority in Phase~4 ensures that they all carry the same decision value.

%%%%%%%%%%%%%%%%%
%  conclusions  %
%%%%%%%%%%%%%%%%%

\section{Conclusions}
\label{sec:conclusions}

We present an \unrel loop construct with \weakR global escape for fault-tolerant multiparty session types (FTMPST) for systems that may suffer from message loss or crash failures.
We prove subject reduction and progress and present a small but relevant case study.

Currently we require all actions within loop programs/bodies to be \unrel.
This ensures that a communication partner is not blocked if a loop is terminated.
An interesting question for further work is how to relax this requirement.
For instance, we may allow for a variant of \weakR branching within loop programs/bodies, where moving to the default branch is not only allowed if the sender is suspected to be crashed but also if the receiver suspects that the sender already terminated its loop or at least already moved to another loop iteration.

Moreover, there are a couple of open problems from \cite{petersNestmannWagner22, petersNestmannWagner23}.
A really difficult challenge is to extend branching to at least some kind of message loss, while maintaining the strong properties of the type system and ensuring that no two alive processes move to different branches.

We also want to study whether and in how far we can introduce \weakR or \unrel session delegation.
Similarly, we want to study \unrel variants of session initialisation including process crashes and lost messages during session initialisation.
\Unrel variants of session initialisation open a new perspective on \MPST-frameworks such as \cite{char16} with dynamically changing network topologies and sessions for that the number of roles is determined at run-time.

As in \cite{petersNestmannWagner23} we fix one set of conditions on failure patterns to prove subject reduction and progress.
We can also think of other sets of conditions.
As already mentioned, we can improve Condition~\ref{cond:all}.\ref{cond:fpDropA} by explicitly using the $ \Expr[id] $ of loops in unreliable messages.
We can also use the failure pattern $ \fpUGet $ to reject the reception of outdated messages.
Therefore, we drop Condition~\ref{cond:all}.\ref{cond:ugetValid} and instead require for each message $ m $ whose reception is refused that $ \fpML $ ensures that $ m $ is eventually dropped from the respective queue and that $ \fpUSkip $ allows to skip the reception of these messages.
An interesting question is to find minimal requirements and minimal sets of conditions that allow to prove correctness in general.

It would be nice to also fully automate the remaining proofs for the distributed algorithm in Section~\ref{sec:example}, namely for validity, agreement, and termination. The approach in \cite{petersWagnerNestmann19} sequentialises well-typed systems and gives the much simpler remaining verification problem to a model checker. Interestingly, the main challenges to adopt this approach are not the \unrel or \weakR prefixes but the failure patterns.

%%%%%%%%%%%%%%%%%%
%  bibliography  %
%%%%%%%%%%%%%%%%%%

\bibliographystyle{eptcs}
\bibliography{ftmpst-loop}

\end{document}